\def\of{\begin{equation}}
\def\zf{\end{equation}}
\def\zff#1{{\label{#1}}\end{equation}}
\def\ofa{\begin{eqnarray}}
\def\zfa{\end{eqnarray}}
\def\zffa#1{{\label{#1}}\end{eqnarray}}
\def\jna#1{(\ref{#1})}			    
\def\jne#1{(\ref{#1})}			    

\def\ch{\hbox{cosh}}
\def\sh{\hbox{sinh}}
\def\cth{\hbox{cotanh}}
\def\th{\hbox{tanh}}
\def\sech{\hbox{sech}}

\def\ni{\noindent}                      
\def\frac#1#2{{#1 \over #2}}            
\def\tr{\hbox{Tr}}			    
\def\Re{\hbox{Re }}			    
\def\Im{\hbox{Im }}			    
\def\tilda#1{{\mathaccent "7E #1}}      

\def\desnoime#1{{\rightline{{\it #1}}\smallskip\hrule\smallskip}}

\newtheorem{defi}{\ni $\bullet$ \underline{{\bf Definition}}}
\newtheorem{lemma}{\ni $\bullet$ \underline{{\bf Lemma}}}

\def\nedelta{\delta\kern-1.0 ex\vrule height 1.3 ex depth-1.14 ex width 0.7 ex \kern 0.15 ex}

\def\babla#1{{}^{\dag}\nabla_{#1} }

\newcount\notenumber
\def\clearnotenumber{\notenumber=0}
\def\note{\advance\notenumber by1 \footnote{$^\the\notenumber}$}
\clearnotenumber

\def\desnoime#1{{}} 
 
\documentstyle[prl,eqsecnum,multicol,aps,fleqn,amsfonts]{revtex}

\begin{document}

\title{Symmetries of Optical Phase Conjugation}
\author {
Predrag L. Stojkov ${}^{\ast}$, 
Milivoj R. Beli{\'c}${}^{\dag}$, 
and \fbox{Marko V. Jari{\'c}${}^\ddag$}}
\address{
${}^{\ast}$ 
UWin! Corporation, GTECH \\
2006 Nooseneck Hill Road, Coventry RI 02816, USA 
\\
${}^\dag$ Institute of Physics
\\
P.O.Box 57, 11001 Belgrade, Yugoslavia
\\ 
${}^{\ddag}$ Department of
Physics, Texas A\&M University
\\ 
College Station TX 77843-4242, USA
\\
}

\maketitle

\begin{abstract} 
Various algebraic structures of 
degenerate four-wave mixing equations of optical phase
conjugation are analyzed. 
Two approaches (the spinorial and the Lax-pair based), 
complementary to each other, 
are utilized for a systematic derivation of conserved quantities.  
Symmetry groups of both the equations and 
the conserved quantities are determined, and the corresponding 
generators are written down explicitly. Relation between these two
symmetry groups is found.
Conserved quantities enable the introduction of new methods for integration of 
the equations in the cases when the coupling $\Gamma$ is either purely real or 
purely imaginary. These methods allow for both geometries of 
the process, namely the transmission and the reflection, 
to be treated on an equal basis. 
One approach to introduction of Hamiltonian and Lagrangian 
structures for the 4WM systems is explored, and the obstacles 
in successful implementation of that programe are identified.
In case of real coupling these obstacles are removable, and 
full Hamiltonian and Lagrangian formulations of the initial system 
are possible.
\end{abstract}


\begin{multicols}{2}
\narrowtext

\section{Introduction}

\desnoime{intro.tex} 

{\it There is a short story prefacing the paper. The work on symmetries in
optical phase conjugation started in early nineties by Predrag Stojkov and
Milivoj Beli\'c. It was interrupted by Predrag's leaving for America in 1992. During the 
stay of M. Beli\'c at the Texas A\&M University in 1995 and 1996, the problem 
and an early draft of the paper were brought to Marko's attention. 
At the time he was phasing
out of quasicrystals, and was open to new ideas. Marko liked the problem and 
agreed to participate. He read the 
manuscript, made numerous changes, and suggested a new direction to it. 
Owing to his commitments for the sabbatical at Cornell University in 1995 and the
visit to Israel in 1996, it was decided to postpone the serious work after he is back.
However, during the Israeli visit Marko was diagnosed with the brain tumor. 
The paper is left essentially unchanged. It is dedicated to his memory.}\\

Steady-state four-wave mixing (4WM) equations describing optical phase
conjugation (OPC) in photorefractive (PR) crystals have been solved up to
now in a number of ways \cite{cronin84,belic85,bledowski87,zozulya88}. A
common feature of all solution methods is that, first, conserved
quantities are determined, and then the number of equations is reduced. 
However, the determination of conserved quantities and the reduction of
equations is usually performed in an ad hoc manner. Furthermore, the solution
of the OPC equations in the two basic geometries of the process, the
transmission geometry and the reflection
geometry, is usually obtained using unrelated methods. 

Apparent symmetries of wave equations have not been used up to
now \cite{fish92,stojkov92} to facilitate the analysis and the
solution of the problem. In this work the symmetries of
the equations and the integrals of motion are investigated  and used 
to present a unified method for systematic derivation of conserved quantities, an 
equal treatment of both geometries, and the reduction in the number of
independent variables \cite{belic94}. Such an analysis allows for not
only an easier handling of
otherwise cumbersome and unrelated relations, but also for a deeper
understanding of the physics of the process. Also, rudiments of
a formal presentation of the problem along the lines of the theory of dynamical
systems are presented.

The geometry of the process is simple. Three laser beams intersect within a 
piece of the PR crystal: two counterpropagating laser pumps
$A_1$ and $A_2$, and a signal $A_4$. Owing to the PR effect, a
fourth wave $A_3$ is generated inside the crystal, that counterpropagates to, 
and is the phase conjugate replica of the signal. There are two main 
channels along which the generation  may proceed.  In the first one, 
the signal wave builds a diffraction grating  with
the pump $A_1$. The other pump is diffracted off that grating and transmitted
across the crystal 
into the PC wave $A_3$. This is the so-called transmission geometry
(TG) of the process.

In the second channel the  signal interferes with the pump
$A_2$, and the beam $A_1$ is reflected off the grating into the PC wave. 
This is the
reflection geometry (RG) of the 4WM process. It is assumed that all waves  
oscillate at the same frequency (the degenerate 4WM). Also, a steady-state
is assumed, and all beams are approximated by plane waves.

The equations of  interest are  the  slowly
varying  envelope wave equations describing 4WM in PR media
\cite{cronin84}. In TG, they are of the form
\vspace{0ex}
\ofa
I A'_1 &=& \Gamma Q_T A_4 , \cr
I A'_2 &=& \bar{\Gamma} \bar{Q}_T A_3 , \cr
I A'_3 &=& - \bar{\Gamma} \bar{Q}_T A_2 , \cr
I A'_4 &=& - \Gamma Q_T A_1 ,
\zfa

\noindent where $I=\sum_{i=1}^{4} | A_i |^2$  is the total
intensity, $\Gamma$ is the coupling constant (complex in general,
but often real in PR media), $ Q_T =  A_1 \bar{A}_4 + \bar{A}_2 A_3 $ represents
the  diffraction  grating amplitude for TG, the prime denotes
spatial derivative along  the propagation $z$ direction, and the bar
denotes complex  conjugation.

In RG, the equations are given by
\vspace{0ex}
\ofa
    I A'_1 &=& - \Gamma Q_R A_3 , \cr
    I A'_2 &=& - \bar{\Gamma} \bar{Q}_R A_4 , \cr
    I A'_3 &=& - \bar{\Gamma} \bar{Q}_R A_1 , \cr
    I A'_4 &=& - \Gamma Q_R A_2 ,
\zfa

\noindent where the RG grating amplitude is given by 
$Q_R = A_1 \bar{A}_3 + \bar{A}_2A_4 $.

In this paper  both geometries are treated on an equal footing, using a
unified RG-like notation:
\vspace{0ex}
\ofa
B_1 &=& A_1 ,  \cr
B_2 &=& A_2 ,  \cr
B_3 &=& A_4 \Pi_{\sigma} + A_3 \Pi_{- \sigma}  
     = \left\{
	\matrix{ A_4 \ \hbox{in TG}, \cr A_3 \ \hbox{in RG}, } 
	\right.  \cr
B_4 &=& A_3 \Pi_{\sigma} + A_4 \Pi_{- \sigma}  
     = \left\{ \matrix{A_3 \ \hbox{in TG},\cr A_4 \ \hbox{in RG}, } 
	\right. 
\zfa

\noindent 
where  $\sigma$ is the switching variable, that has value $+1$ for TG and
$-1$ for RG, and  $ \Pi_{\pm\sigma} = ( 1 \pm \sigma) /2$ are the
corresponding "projectors."

The "equations of motion" (EOM) are now
\vspace{0ex}
\ofa
I B'_1 &=&  \sigma \Gamma Q B_3 , \cr
I B'_2 &=&  \sigma \bar{\Gamma} \bar{Q} B_4 , \cr
I B'_3 &=& - \bar{\Gamma} \bar{Q} B_1  , \cr
I B'_4 &=& - \Gamma Q B_2  ,
\zffa{eom}

\noindent where the intensity is given by 
$I=\sum_{i=1}^{4} |B_i|^2$  and the grating 
amplitude by $ Q = B_1 \bar{B}_3 + \bar{B}_2 B_4$.

The analysis is organized as follows. Two methods to 
derive the integrals of motion (IOM) are discussed in 
Section \ref{section_on_ioms}. The first method 
is based on the observation that 4WM EOM have a special symmetric 
form that is allowing an equivalent {\it spinorial} formulation. Such a form 
of EOM leads directly to the derivation of the full set of "regular" IOM
as suitable bi-spinorial combinations. The symmetries of these IOM are the 
special unitary groups: $SU(2)$ for TG, and 
$SU(1,1)$ for RG. Initial spinor-like doublets of fields turn out
to transform as the fundamental irreducible representations of these groups, 
thus justifying the name "spinors". It is indicated how they can be used to 
reduce the number of dynamical variables. 

In the second method the Lax pair approach is utilized. The diadic products of 
the 4WM spinors are 
used as possible choices for the evolving member (${\cal L}$) of the Lax pair 
problem. The traces of products of these matrices represent IOM. It was 
established that all higher order IOM are various combinations of the basic 
IOM already obtained by the spinorial approach. At the end of Section 
\ref{section_on_ioms} two special cases ($\Gamma\in{\Bbb R}$ and 
$\Gamma\in i{\Bbb R}$) are considered in some detail.

In Section \ref{section_on_eoms} attention is focused 
on the derivation of the symmetry groups of EOM, 
and the corresponding generators. The relation between these
symmetries and the symmetries of IOM is discussed.

In Section \ref{section_on_sols} the symmetries of IOM are used to
write the solutions of EOM (for $\Gamma\in{\Bbb R}$) in terms of
elementary 
transcendental functions. Then an alternative solution procedure is explored. 
In the last part of Section \ref{section_on_sols} the $\Gamma\in i{\Bbb R}$ case is solved completely.

The possibility of introducing the Hamiltonian and Lagrangian
description of 4WM EOM is explored 
in Section \ref{section_on_pseudo}. 
Section \ref{section_with_conclusions} offers some conclusions and identifies
open questions for future research.

\section{Integrals of motion and their symmetries}

\label{section_on_ioms}

\subsection{Preliminaries}

\desnoime{iom\_prelim.tex}

In this work the 4WM equations are treated as a dynamical system
defined on the phase space $V:\equiv \tilda{V}/\rho
\equiv {\Bbb R}^8$, 
with the time variable $z$. Here $\tilda{V}$ is the full 
sixteen-dimensional space $\tilda{V}:\equiv{\Bbb C}^8\equiv {\Bbb R}^{16}$ 
with the complex coordinates $\{x^\mu\}\equiv\{B_i,\bar{B}_i\}$ 
and $\rho$ is the equivalence relation (analyticity condition) satisfied 
by the 4WM system: $x^{i+4}=(x^i)^*$ 
[i.e. $\bar{B}_i = (B_i)^*$] for $i=\overline{1,4}$. 
The tangent space $TV$ (the space of vector fields on $V$),
is spanned by the {\it coordinate} basis 
$\{ \partial_{\mu}\}\equiv\{
\partial_i = {\partial}/{\partial B_i} ; \bar{\partial}_{i} =
{\partial}/{\partial \bar{B}_i}\} $. 
The dynamics on the space $V$ is given by the
trajectory $ {\it c}:{\Bbb R}_z \rightarrow V$. It is
described by the velocity vector-field $\vec{F}\in TV$:
\vspace{0ex}
\ofa
\vec{F} = 
\mu (\sigma B_3 \partial_1 - B_2
\partial_4) + \bar{\mu} (\sigma B_4 \partial_2 - B_1
\partial_3) +  c.c.
\cr
&&
\zfa
\noindent where $\mu:\equiv {\Gamma Q}/{I}$.
For a general function $f(z,x) \in C^1({\Bbb R}_z\times V)$,
the corresponding evolution equation is
\vspace{0ex}
\of 
\left(\frac{d f}{d z}\right) = {(\partial_{z} + \vec{F} ) f } . 
\zf

In general, an integral of motion (IOM) $q(z,x)$ is a function that is
constant along the trajectory $c$, i.e. ${(\partial_z +\vec{F})} q |_c = 0$
 ({\it on-shell constancy}). 
Here the more restrictive definition of IOM is used:
instead of an on-shell constancy, the condition of 
{\it off-shell} constancy 
[$(\partial_z+\vec{F}) q = 0$ in whole $V$] is used.
Also, only the  integrals $q(x)$ 
without the explicit time-dependence are considered, leading to the 
defining equation 
\vspace{0ex}
\of
\vec{F}{q}=0 .
\zff{konstanta}

There is no general procedure for finding IOM, and one has to take
into account various specifics of the system at hand. For example,
one may resort to the brute-force solution procedure, which is
based on the observation that $\vec{F}$ is a linear differential 
operator of the zeroth degree of homogeneity 
(i.e. upon its action on some homogeneous function 
$P_s(x)$ of degree $s$, it produces another homogeneous function of 
the same degree $s$). This allows one to replace Eq. \jna{konstanta} 
by the set of infinitely many equations:
\vspace{0ex}
\of
\vec{F} q_{(s)} = 0 ,
\zff{punokonstanti}

\noindent for $s \in {\Bbb N}$, 
where $q_{(s)}$ are the components of $q$ with the fixed
degree of homogeneity $s$ ($q = \sum_{s=1}^\infty q_{(s)}$).
In general, Eqs. \jna{punokonstanti} can be solved by a general
ansatz (the summation over the repeated indices is assumed):
\vspace{0ex}
\ofa
q_{(1)} &=& \alpha_{\mu} x^{\mu}  , 
\cr
q_{(2)} &=& \alpha_{\mu\nu} x^{\mu} x^{\nu} , 
\cr 
&&
\cdots
\zfa
\noindent which turns Eq. \jna{punokonstanti} into a set of 
conditions for the matrices $\alpha$. After some algebra, 
one finds that there are no integrals of the first degree, 
and that there are several of the second and higher degrees.

\subsection{Spinorial formalism}

\label{subsection_on_spinors}

\desnoime{iom\_spinors.tex}

There exists a more elegant way to find integrals
of motion of the second degree of homogeneity in the 4WM equations 
\cite{stojkov92}. 
It is based on the
fact that convenient pairs of columns ("spinors") can be formed:
\vspace{0ex}
\of 
\left|\psi_1\right>
= 
\left({B_1 \atop B_3}\right) , \ \ 
\left|\psi_2\right> 
=
\left({B_4 \atop - \sigma B_2}\right).
\zf
Now the equations of motion \jna{eom} can  be written as
a pair of matrix equations:
\vspace{0ex}
\of
\left|\psi_j\right>'  =  {\bf m}    \left|\psi_j\right>, \  \ (j=1,2)
\zff{dirac}

\noindent 
where the "evolution matrix"  
${\bf m}$ is
\vspace{0ex}
\of 
{\bf m} = \pmatrix{ 0 &  \sigma \mu \cr - \bar{\mu} & 0},
\zf

\noindent and $\mu=\Gamma Q/I$. 
The  matrix ${\bf m}$ is traceless and Hermitian (for TG) 
or skew-Hermitian (for RG), so it belongs to the $su(2)$ algebra for TG,
or to the $su(1,1)$ for RG.

IOM are found using a simple Lemma:

\medskip\hrule\smallskip
{\lemma : 
A pair of linear matrix equations
\vspace{0ex}
\of 
\left|\psi_A\right>' = {\bf m}_{A} \left|\psi_A\right>  , \
\left|\psi_B\right>'  = {\bf m}_{B} \left|\psi_B\right>  ,
\zf
\noindent has an integral of motion 
$\left<\psi_A\right| {\bf n}\left|\psi_B\right>$,
if there exists a constant matrix ${\bf n}$ such that
\vspace{0ex}
\of 
{\bf n} {\bf m}_B  +  {{\bf m}^ {\dag}}_A
{\bf n} = 0,
\zff{definisuca0}

\noindent where the dagger denotes the adjoint matrix.
}
\smallskip\hrule\medskip

In 4WM the IOM are searched for in two possible forms, as
$\left<\psi_i\right| {\bf n}\left|\psi_j\right>$ or as $\left<\bar{\psi}_i\right|
{\bf n}\left|\psi_j\right>$. For the first form, we have the defining
equation \jna{definisuca0} specified as
\vspace{0ex}
\of 
{\bf n}{\bf m}+{\bf m}^{\dag}{\bf n}=0,
\zff{definisuca1}
\noindent whereas for the second form of integrals, the defining
equation is
\vspace{0ex}
\of 
{\bf n}{\bf m} + {\bf m}^{T}{\bf n} = 0.
\zff{definisuca2}
For the general (complex) $\Gamma$, the unique solutions 
(up to rescaling by a constant) of these
defining equations are the matrices
\vspace{0ex}
\of 
{\bf n}_1  
=  \pmatrix{ 1 & 0 \cr  0 &  \sigma}
= \left\{ 
\matrix{
{\bf 1} \ \hbox{for TG}, \cr 
{\bf \sigma}_3 \ \hbox{ for RG}, 
}  
\right.
\zf

\noindent for \jna{definisuca1} and
\vspace{0ex}
\ofa 
{\bf n}_2  
&=&  
\pmatrix{ 0 & - \sigma \cr \sigma  & 0} 
=  - \sigma  i {\bf \sigma}_2 =
\left\{
\matrix{-i {\bf \sigma}_2 \ \hbox{for TG},
\cr i {\bf \sigma}_2 \
\hbox{ for RG},
}
\right. 
\cr
&&
\zfa
\noindent for \jna{definisuca2}, 
where ${\bf \sigma}_{j}$  are the Pauli matrices.
The corresponding integrals are
\vspace{0ex}
\ofa 
q_1  &=& \left<\psi_1\right|  {\bf n}_1 \left|\psi_1\right> = I_1  + \sigma I_3 
 , \cr 
q_2  &=& \left<\psi_2\right|  {\bf n}_1 \left|\psi_2\right> = I_2 + \sigma I_4 
 , \cr 
q_3  &=& \left<\psi_2\right| {\bf n}_1 \left|\psi_1\right> = B_1\bar{B}_4 - B_3\bar{B}_2
 , \cr 
q_4  &=& \left<\bar{\psi}_1\right| {\bf n}_2 \left|\psi_2\right> = B_1 B_2 + \sigma B_3 B_4
 . 
\zfa

\ni Since these IOM are present for arbitrary complex coupling
$\Gamma$, they are said to be the {\it regular} IOM of the 4WM system.
Later it will be shown that for special choices of $\Gamma$ this 
system possesses additional ({\it exceptional}) IOM.
 
Not all of the conserved quantities $q_1$,  $q_2$ , $q_3$,
$\bar{q}_3$ , $q_4$ and $\bar{q}_4$  are independent.  
There exists a relation
\vspace{0ex}
\of 
|q_4 |^2  +  \sigma |q_3 |^2  = q_1 q_2 
\zff{q1q2}

\noindent that reduces the number of (real) integrals of motion to
five. Using the integrals, one can express the conjugated fields as
dependent variables:
\begin{eqnarray}
\bar{B}_1 &=& [ \sigma B_3 \bar{q}_3 + B_2 q_1 ]/q_4  , \cr
\bar{B}_2 &=& [B_1 q_2 - \sigma B_4 q_3 ]/q_4  , \cr 
\bar{B}_3 &=& [ B_4 q_1 - B_1\bar{q}_3 ]/q_4  , \cr 
\bar{B}_4 &=& [ B_2 q_3 + B_3 q_2 ]/q_4  . 
\end{eqnarray}

A more natural way to reduce the number of variables using
conserved quantities is to introduce {\it polar} coordinates,
suggested by the form of the conserved quantities $q_1$  and $q_2$:
\vspace{0ex}
\ofa 
B_1  = \sqrt{q_1} c(\sigma,\alpha_1 )\exp(i\beta_1 )
 , \cr
B_2  = \sqrt{q_2} c(\sigma,\alpha_2 )\exp(i\beta_2 )
 , \cr
B_3  = \sqrt{q_1} s(\sigma,\alpha_1 )\exp(i\gamma_1 )
 , \cr
B_4  = \sqrt{q_2} s(\sigma,\alpha_2 )\exp(i\gamma_2 )
 .
\zffa{polarne}

\noindent Here the new variables are the six angles: $\alpha_1$,
$\alpha_2$, $\beta_1$, $\beta_2$, $\gamma_1$, and $\gamma_2$. The
symbols  ${c}$ and ${s}$ stand for the trigonometric
cosine and sine functions in the TG case, and for the hyperbolic
cosine and sine in RG case (see Appendix A).

To further ilucidate the connection between IOM and the new
variables, it is convenient to employ the integral $q$, introduced below [Eq. 
\jna{ku}]. It 
can be expressed in terms of the new variables, and the result is
\vspace{0ex}
\ofa 
q  &=&  q_1^2   +  q_2^2    +  2 q_1 q_2  [ c(\sigma,2\alpha_1
) c(\sigma,2\alpha_2 ) + 
\cr 
&& 
+  \sigma   s(\sigma,2\alpha_1 ) s(\sigma,2\alpha_2
)\cos(\Phi)]
 ,
\zfa

\noindent where $\Phi=\beta_1 +\beta_2 -\gamma_1 -\gamma_2$  is
the so-called relative phase.  From  the  spherical  and  the
hyperbolic \cite{berger77} trigonometry it is known
that the expression in brackets can be understood as a cosine
(hyperbolic cosine) of some angle  $\rho$,  so  that $2\alpha_1$ ,
$2\alpha_2$ and $\rho$ are the sides of a spherical (hyperbolic)
triangle,  and $\Phi$ is its central angle. 
Therefore
\vspace{0ex}
\of
c(\sigma,\rho) = {q-q_1^2-q_2^2\over 2 q_1 q_2} = {const.}
\zf

\noindent and $\Phi$ only depends on $\alpha_1$ and $\alpha_2$ :
\vspace{0ex}
\of
\cos(\Phi) =  { c(\sigma,\rho) - c(\sigma,2\alpha_1) 
c(\sigma,2\alpha_2) \over
\sigma s(\sigma,2\alpha_1) s(\sigma,2\alpha_2)} .
\zf

It is easy to check that:
\vspace{0ex}
\ofa 
|q_4 |^2  &=&  q_1 q_2  c\left(\sigma,{\rho\over 2} \right)^2, 
\cr 
|q_3 |^2  &=&  q_1 q_2  s\left(\sigma,{\rho\over 2} \right)^2, 
\zfa

\noindent in agreement with Eq. \jna{q1q2}. Thus, there are five
independent  real  conserved quantities: $q_1$,  $q_2$, $\rho$, and
the phases of $q_3$ and $q_4$. The solution of EOM using these quantities
is performed in Sec. \ref{section_on_sols}.

\subsection{ Lax Pairs}
\label{subsection_on_lax}
\desnoime{iom\_lax.tex}

The Lax pair representation (if it exists) helps determination 
of the integrals of motion. In general, 
if given dynamical system admits a Lax pair representation
\vspace{-1ex}
\of
\frac{d\hat{\cal L}}{dz}=[\hat{\cal M},\hat{\cal L}] ,
\zff{lax}
\noindent 
where $\hat{\cal L}$ and $\hat{\cal M}$ are suitably chosen operators or
matrices, then all the traces  
$Tr(\hat{\cal L}^k)$ ($k\in{\Bbb N}$) 
are IOM. The determination of such a Lax pair of operators 
$(\hat{\cal M}, \hat{\cal L})$ 
is usually the hardest part of the problem. The brackets in Eq. (\ref{lax})
stand for the {\it commutator}.

In the case of 4WM, the suitable matrices are easy to find,
starting from the compact form of the spinorial EOM \jna{dirac}
and their conjugated equations: 
\vspace{-1ex}
\ofa
\partial_z |\psi_\mu\rangle &=& {\bf m} |\psi_\mu\rangle, \cr
\partial_z \langle\psi_\mu| &=& -\sigma  \langle\psi_\mu| {\bf m}, 
\zfa
\ni where the index $\mu=i,\bar{i}$ and 
$|\psi_{\bar{i}}\rangle :\equiv {\bf n}_1 {\bf n}_2 |\bar{\psi}_i\rangle$. 
The following matrices
\vspace{-1ex}
\of
{\bf \cal L}_{\mu\nu} :\equiv |\psi_\mu\rangle\langle\psi_\nu|  {\bf n}_ 1  , 
\zf
\noindent  satisfy the Lax pair equations
\vspace{-1ex}
\of 
\partial_z {\bf \cal L}_{\mu\nu}
= 
[ {\bf m}, {\bf \cal L}_{\mu\nu}].
\zf
The corresponding Laxian IOM are
given by
\vspace{-1ex}
\of
q_{\mu_1\nu_1\cdots\mu_k\nu_k} 
:\equiv 
\tr\left({\bf \cal L}_{\mu_1\nu_1} \cdots
{\bf \cal L}_{\mu_k\nu_k}\right).
\zf
For $k=1$ we have
\vspace{-1ex}
\of
\left(q_{(1)\mu\nu}\right)
=
\pmatrix{
q_1&q_3&0&-q_4\cr
\bar{q}_3&q_2&-q_4&0\cr
0&-\bar{q}_4&\sigma q_1&\sigma \bar{q}_3\cr
-\bar{q}_4&0&\sigma q_3&q_2}.
\zf
For higher $k$, the resulting IOM are the products of $q_{(1)}$. 
For example,
\vspace{-1ex}
$$q_{(2)\mu\nu\alpha\beta} = q_{(1)\alpha\nu} q_{(1)\mu\beta}. $$
Thus, the higher Laxian IOM are not yielding any new independent integrals.

An alternative variant of the Lax pair approach is presented in  
Appendix \ref{app_lax}.

\subsection{I-Symmetry}

\desnoime{iom\_isymm.tex}

{\defi  :
The symmetry of the set of
integrals $\{q_\alpha\}$ ({\bf the I-symmetry}) 
is the mapping $\{x^\mu\}$
$\rightarrow$ $\{x'^\mu\}$ which preserves the analytical
structure $(x')^{i+4}=((x')^i)^*$ and leaves all the integrals
invariant $q_\alpha(x') =q_\alpha(x)$. 
}

\smallskip\hrule\medskip

I-symmetries define the algebraic structure of the sistem at hand.
In practice one first calculates the {\it infinitesimal} I-symmetries, 
given by
\vspace{0ex}
\of 
\delta q_\alpha(x) 
\equiv 
\vec{l} q_\alpha 
\equiv
\delta x^{\mu} \partial_{\mu} q_\alpha = 0 , 
\zff{master1}

\noindent where $\vec{l}=\omega^\mu(x)\partial_\mu$
is the generating vector field, 
and then establishes the {\it large} 
(non-infinitesimal) I-symmetries, by exponentiating the
infinitesimal ones. This is the standard procedure in the theory 
of Lie-groups.

Although in general the coefficients $\omega^{\mu}$ are nonlinear
functions of $\{x\}$ (the nonlinear I-symmetries), here
only the linear I-symmetry algebras will be considered. 
These can be calculated easily by a linear ansatz
$\omega^{\mu} = a^{\mu}{}_{\nu} x^{\nu}$. In this way 
a general linear symmetry 
of the full set $\{q_1,\cdots,\bar{q}_4\}$ of the regular IOM 
is found:
\vspace{0ex}
\ofa 
2\delta B_1 
&=&   
+ i {\epsilon_3}B_1 
+ {(\epsilon_2 + i \epsilon_1 )} B_3 
 , \cr 
2\delta B_2 
&=& 
- i {\epsilon_3}B_2 
+ {(\epsilon_2  - i \epsilon_1  )} B_4 
 , \cr 
2\delta B_3 
&=& 
- i {\epsilon_3}B_3 
-  \sigma {(\epsilon_2  - i \epsilon_1 )} B_1 
 , \cr
2\delta B_4 
&=&   
+ i {\epsilon_3}B_4 
- \sigma {(\epsilon_2 + i \epsilon_1  )} B_2
. 
\zffa{raspisano}

In the spinor notation the matrix form of I-symmetries is
\vspace{0ex}
\of
\delta \left|\psi_{1,2}\right> = {\bf \Sigma}^{T} \left|\psi_{1,2}\right> ,
\zff{malispinor}

\noindent where the traceless matrix ${\bf \Sigma}$ is given by
\vspace{0ex}
\of 
{\bf \Sigma} 
= 
\frac{1}{2} 
\pmatrix{ 
i {\epsilon_3}
& 
-\sigma (\epsilon_2  - i \epsilon_1)  
\cr 
\epsilon_2 + i \epsilon_1
& 
- i {\epsilon_3}
}
= 
i \epsilon_a {\bf S}_a
 ,
\zf

\noindent and $\{\epsilon_a\}$ are real parameters. The basis
matrices (the {\it generators}) are
\vspace{0ex}
\ofa
{\bf S}_1 
&=& 
{1\over 2}\left(\matrix{0&\sigma\cr 1&0}\right) 
= 
\left\{
\matrix{
{1\over 2}{\bf \sigma}_1 \ \ \hbox{for TG},
\cr 
-{i\over 2}{\bf \sigma}_2 \ \ \hbox{for RG},}
\right.  
\cr
{\bf S}_2 
&=& 
{1\over 2}\left(\matrix{0&\sigma i\cr -i&0}\right) 
=
\left\{
\matrix{
-{1\over 2}{\bf \sigma}_2 \ \ \hbox{for TG},
\cr 
-{i\over 2}{\bf \sigma}_1 \ \ \hbox{for RG},}
\right.  
\cr
{\bf S}_3
&=& 
{1\over 2}{\bf \sigma}_3
.
\zfa

The set of all matrices ${\bf \Sigma}$ that are traceless and
satisfy the generalized hermiticity condition
${\bf \Sigma}^{\dagger}{\bf \eta} + {\bf \eta}{\bf \Sigma} = 0$
forms the Lie-algebra $su(2)$ for TG, and $su(1,1)$ for
RG. The matrix ${\bf \eta}=  {\bf n}_1$ 
is called the $su(2)/su(1,1)$ {\it metric} matrix.
The generators $\{{\bf S}_a\}$ obey the standard commutation
relations
\vspace{0ex}
\ofa
[{\bf S}_1,{\bf S}_2] 
&=& 
- i  \sigma {\bf S}_3 \
, 
\cr
[{\bf S}_2,{\bf S}_3] 
&=& 
- i  {\bf S}_1 \ 
,
\cr
[{\bf S}_3,{\bf S}_1] 
&=& 
- i  {\bf S}_2 \ .
\zfa
Thus, both $\left|\psi_1\right>$ and $\left|\psi_2\right>$ are transforming
according to the {\it fundamental} (spinorial) representation of the
corresponding algebra ${\it g}_I$.

Every finite Lie-algebra {\it g} has the corresponding 
Lie-group ${\cal G}$ of
"large" transformations, obtained via exponential 
mapping: 
$$\forall {\bf \Sigma} \in {\it g} \Rightarrow 
{\bf G} :\equiv \exp(i{\bf \Sigma}) \in {\cal G}. $$ 
For ${\it g}_I= su(2)$ (the TG case), the group is ${\cal G}_I =
SU(2)$, and for ${\it g}_I= su(1,1)$ (the RG case), 
the group is $SU(1,1)$,
the noncompact version of $SU(2)$. Both groups can be represented
by sets of $2\times 2$ complex matrices ${\bf G}$ that are
unimodular ($\det{\bf G}=1$) and (pseudo)unitary
(${\bf G}^{\dag}{\bf \eta}{\bf G}={\bf \eta}$). 

The Cayley-Klein 
parameterization of the general $SU(2)/SU(1,1)$ group element 
\vspace{0ex}
\of {\bf G} = 
\pmatrix{y_1 + i y_2 & y_3 + i y_4 \cr -
\sigma (y_3 - i y_4) & y_1 - i y_2 }, \zf
\noindent  where $y_{1,\cdots,4} \in
{\Bbb R}$, turns the unimodality condition into a geometric relation
(the definition of the parameter manifold of the group ${\cal G}_I$):
\vspace{0ex}
\ofa \det{{\bf G}} &=& (y_1)^2 +(y_2)^2+\sigma \left[(y_3)^2 +(y_4)^2\right]
= 1.\cr &&\zfa
\noindent Thus, the parameter {\it manifold} for $SU(2)_I$ is the
sphere ${\Bbb S}^3$, and for $SU(1,1)_I$ it is the 
hyperboloid ${\Bbb H}^3$, both embedded in ${\Bbb R}^4$.
(For a short classification of hyperboloids in 
${\Bbb R}^4$ see Appendix \ref{appendix_on_hyperboloids}.)

From this fact alone, one could 
expect that the TG case will be expressed in a natural way 
in terms of the trigonometric functions, and the RG case in terms of
both the trigonometric functions (compact dimensions) and the
hyperbolic functions (noncompact dimensions). In this sense the
cases are "twins", i.e. there is a number of equations
holding in both cases, up to the exchange of the 
trigonometric/hyperbolic functions.

\subsection{Action of I-Symmetries on the Lax variables}

\label{isymm_acting_on_lax}

\desnoime{iom\_isymm\_lax.tex}

It is of interest to know the action of the 
I-symmetries on the Lax matrices 
${\bf \cal L}_{\mu\nu}$. 
Since the Lax  matrices are constructed out of 
the basic spinors, some regularity must be induced in 
the transformation law of these variables.

For example, 
for ${\bf \cal L}_{11}=|\psi_1\rangle \langle \psi_1|{\bf n}_1$,
the action of the infinitesimal I-symmetry yields
$$\delta {\bf \cal L}_{11} =
{\bf \Sigma}^T  {\bf \cal L}_{11} +
{\bf \cal L}_{11} {\bf n}_1^{-1} {\bf \Sigma}^* {\bf n}_1.$$
\ni Owing to matrix identities 
${\bf n}_1^{-1}={\bf n}_1$ and ${\bf n}_1 {\bf \Sigma}^* {\bf n}_1
= - {\bf \Sigma}^T$, this expression simplifies to 
\vspace{0ex}
\of
\delta {\bf \cal L}_{11} 
=
\left[
{\bf \Sigma}^T , {\bf \cal L}_{11}
\right].
\zf

\ni This represents the {\it adjoint action} 
of the I-symmetry on 
${\bf \cal L}_{11}$. In a similar way, one finds 
that the same transformation law is valid for all ${\bf \cal L}_{\mu\nu}$:
\vspace{0ex}
\ofa
\delta {\bf \cal L}_{\mu\nu} &=& 
\left[{\bf \Sigma}^T,{\bf \cal L}_{\mu\nu}\right].
\zfa
Thus, due to cyclic invariance of the matrix trace operation, 
all Laxian IOM are invariant upon the action of I-symmetries. This is expected.

\subsection{Exceptional IOM}

\desnoime{iom\_except\_intro.tex}

The "regular" IOM, obtained in  the subsection \ref{subsection_on_spinors},
form the full set of IOM for the complex coupling $\Gamma$. 
However, in the special cases when $\Gamma$ is either real or imaginary, 
there exist additional IOM. These will be called the "exceptional" IOM. 

To see the significance of these special cases, let us evaluate the "time"-change of 
the grating amplitude $Q$:
\vspace{0ex}
\of 
I Q'    = - \Gamma Q q_5.
\zff{opa1}

\ni Here $q_5$ is the expression  $q_5  = I_1  + I_2  -  \sigma (I_3  + I_4)$, 
whose "time"-change is 
\vspace{0ex}
\of 
I q'_5  = 4 \sigma \Re(\Gamma)|Q |^2. 
\zff{opa2}

\ni Notice that $q_5$
is IOM in the case of imaginary $\Gamma$ (so, it is an 
"exceptional" IOM). However, when  $\Gamma$ is  a
complex number, 
this quantity turns out to be a suitable variable for later calculations.

From equations \jna{opa1} and \jna{opa2} another conserved quantity 
(for the general, complex $\Gamma$) is
obtained:
\vspace{0ex}
\of 
q = q^2_5  + 4  \sigma |Q |^2
 .
\zff{ku}

\noindent This quantity is IOM of the fourth order. It
does not  carry  any independent information, since it is
reducible to the already known regular IOM:
\vspace{0ex}
\of 
q =  (q_1  + q_2)^2    - 4 \sigma |q_3 |^2 
 . 
\zff{ku1}

\noindent Nevertheless, $q_5$ plays an important role in 
one of the two presented procedures for solving EOM in the $\Gamma\in{\Bbb R}$ case. 

From Eq. \jna{opa1}, two important relations follow:
\vspace{0ex}
\ofa
I |Q|'       &=& - \Re(\Gamma) |Q| q_5
 ,\cr
I {\arg(Q)}' &=& - \Im(\Gamma) q_5
 .
\zffa{vazna}

\ni These equations indicate the existence of two important special cases: 
 $\Gamma \in {\Bbb R}$ and $\Gamma \in i{\Bbb R}$.  
The case $\Im\Gamma=0$ implies $\arg Q =
{const}$,  so that $\phi:\equiv\arg\mu={const}$,
while the case  $\Re\Gamma=0$ implies that 
$|Q|={const}$. 
The $\Gamma\in{\Bbb R}$ case is considered first.

\subsubsection{ $\Gamma\in{\Bbb R}$}
\desnoime{iom\_except\_R.tex}

Analysis of this case is based on the fact that the phase $\phi$
of the grating amplitude $Q$ is constant for real couplings. This
allows introduction of a new independent variable 
$\theta(z)=\int_0^z dz' |\mu(z')| + \theta_0$, which 
casts the problem into a linear form.  
The matrix  ${\bf m}$ is now replaced by
\vspace{0ex}
\of 
{\bf \tilda{m}} 
=
\pmatrix{0&\sigma\nu \cr-\bar{\nu}&0}
 , 
\zf
\noindent where $\nu:\equiv \exp({i\phi})=const.$

The defining relation \jna{definisuca1} has as solutions not
only the matrix ${\bf n}_1$, but also the new one:
\vspace{0ex}
\of
{\bf{\bf n}}_3\equiv \pmatrix{0&\nu\cr -\bar{\nu}&0}
 ,
\zf
\noindent which is anti-Hermitian 
${\bf n}_3^{\dag}=-{\bf n}_3$.
Along the same lines, the defining relation \jna{definisuca2} has
as solutions both the matrix $ {\bf n}_ 2$ and the new one:
\vspace{0ex}
\of
{\bf{\bf n}}_4\equiv  \pmatrix{\bar{\nu}&0\cr 0&\sigma\nu},
\zf
\noindent which is symmetric.
Having the new matrices $  {\bf{\bf n}}_3$ and $ {\bf n}_ 4$
that satisfy the Lemma, a set of additional
conserved quantities can be constructed:
\vspace{0ex}
\ofa w_1 
&:\equiv& 
\left<\psi_1\right|  {\bf n}_ 3\left|\psi_1\right> 
= 
2 i \Im(\nu \bar{B}_1 B_3) 
\, \cr 
w_2 
&:\equiv& 
\left<\psi_2\right| {\bf n}_3\left|\psi_2\right>
=
- 2 i  \sigma \Im(\nu \bar{B}_4 B_2) 
 , \cr 
w_3
&:\equiv&
\left<\psi_1\right|  {\bf n}_ 3\left|\psi_2\right>
= 
- \sigma \nu \bar{B}_1 B_2 -\bar{\nu} \bar{B}_3 B_4  
\cr 
w_4 
&:\equiv&
\left<\bar{\psi}_1\right|  {\bf n}_ 4\left|\psi_1\right>
=
\bar{\nu} B_1^2  + \sigma \nu B_3^2 
 , \cr 
w_5 
&:\equiv& 
\left<\bar{\psi}_1\right| {\bf n}_4\left|\psi_2\right>
= 
\bar{\nu} B_1 B_4 - \nu B_3 B_2 
 , \cr 
w_6
&:\equiv&
\left<\bar{\psi}_2\right|  {\bf n}_ 4\left|\psi_2\right>
= 
\bar{\nu} B_4^2 + \sigma \nu B_2^2
 . 
\zfa

\noindent Note that $\left<\psi_2\right|{\bf n}_3\left|\psi_1\right> =-w_3$ and $\left<\bar{\psi}_2\right|$ ${\bf n}_ 4$ $\left|\psi_1\right>$ $=$
$w_5$.

One can extend the Lax procedure, in the spirit of the subsection 
\ref{subsection_on_lax}, to this case as well. The Lax matrices are now 
${\bf \cal L}_{\mu\nu}^{({\Bbb R})} :\equiv 
\left|\psi_\mu\left>\right<\psi_\nu\right| {\bf n}_3$, 
and the corresponding Lax equations
\vspace{0ex}
\of
\partial_\theta {\bf \cal L}_{\mu\nu}^{({\Bbb R})} 
= 
[{\bf \tilda{m}}, {\bf \cal L}_{\mu\nu}^{({\Bbb R})}]
.\zf 

\ni The corresponding Laxian IOM of the first order are
\vspace{0ex}
\of
\tr{\bf \cal L}_{\mu\nu}^{({\Bbb R})} = 
\pmatrix{
w_1&-\bar{w}_3&w_4&w_5\cr
w_3&w_2&w_5&w_6\cr
-\bar{w}_4&-\bar{w}_5&-\sigma w_1&-\sigma w_3\cr
-\bar{w}_5&-\bar{w}_6&\sigma \bar{w}_3&-\sigma w_2
}.
\zf

\ni These IOM are the same as the ones already 
obtained through the spinorial approach.
As mentioned, a more general Lax pair procedure is presented in 
Appendix \ref{app_lax}. 

Action of an I-symmetry on ${\bf \cal L}_{\mu\nu}^{({\Bbb R})}$ produces
$$
\delta{\bf \cal L}_{\mu\nu}^{({\Bbb R})}
= 
{\bf \Sigma}^T {\bf \cal L}_{\mu\nu}^{({\Bbb R})}
+
{\bf \cal L}_{\mu\nu}^{({\Bbb R})} {\bf n}_3^{-1} 
{\bf \Sigma}^{T\dagger} {\bf n}_3.
$$

\ni Here the condition ${\bf n}_3^{-1} {\bf \Sigma}^{T\dagger} {\bf n}_3 = - {\bf \Sigma}^T$, 
necessary for the covariant form of action, can be achieved in  
different ways:

$\bullet$ Case 1: $\nu^2=-\sigma$. 
This corresponds to $\phi=\pm \pi/2$ (in TG) and to $\phi=0 \ \hbox{ or } \ \pi$ 
(in RG). This case allows for the full $su_I$ symmetry, i.e. 
all three $\epsilon_a$ parameters can have non-zero values. However, only the diagonal part of ${\bf \Sigma}^T$ figures in the transformation law:
\vspace{0ex}
\of
\delta{\bf \cal L}_{\mu\nu}^{({\Bbb R})}
= \frac{i\epsilon_3}2 [{\bf \sigma}_3,{\bf \cal L}_{\mu\nu}^{({\Bbb R})}].
\zf

$\bullet$ Case 2: $\nu^2 \ne -\sigma$. 
Here only the diagonal part of I-symmetries survives, 
i.e. $\epsilon_1$ and $\epsilon_2$ have to be set equal to zero. 
The transformation law still has the same form as in the case above.

Thus, the $w$ IOM are invariant under the full $su_I$ symmetry algebra 
if $\nu^2=-\sigma$,
and under the $u_1$ subalgebra generated by ${\bf \sigma}_3/2$ 
if $\nu^2\ne-\sigma$.

An important special case is the phase conjugation, when the
relative phase $\Phi$ ($\equiv$ $\beta_1$ $+$ $\beta_2$ $-$
$\gamma_1$ $-$ $\gamma_2$) is constant ($0$ or $\pi$). Then,
using relations \jna{polarne} and the fact that the argument
$\phi$ ($=$ $\beta_1$ $-$ $\gamma_1$ $=$ $\gamma_2$ $-$ $\beta_2$)
of $\mu$ is constant, the following values for the
integrals $w$ are obtained:
\vspace{0ex}
\ofa 
w_1 &=& 0 
 , \cr
w_2 &=& 0 
 , \cr 
w_3 &=& 
-\sqrt{q_1q_2} c(\sigma,\alpha_1-\alpha_2)\exp(i(\beta_2-\gamma_1))
 , \cr 
w_4 &=& 
q_1 \exp(i(\beta_1+\gamma_1))
 , \cr 
w_5 &=&  
\sigma q_2 \exp(i(\beta_2+\gamma_2))
 , \cr 
w_6 &=& 
-\sqrt{q_1q_2} s(\sigma,\alpha_1-\alpha_2)\exp(i(\beta_1 +\beta_2)).
\cr
&& 
\zfa
\ni These relations imply that all the phases $\beta_1$, $\beta_2$, $\gamma_1$, $\gamma_2$ are constant, and that the $\alpha$-variables are
linearly dependent: $\alpha_1-\alpha_2=\hbox{constant}$.
Hence, all the fields essentially depend on only one real quantity, for
example on $\alpha_1$.

\subsubsection{ $\Gamma\in i{\Bbb R}$}

\desnoime{iom\_except\_iR.tex}

The case of $\Gamma$ imaginary has only one exceptional integral, $q_5$. 
The corresponding I-symmetry of the set $\{q_1,...,q_5\}$ is restricted 
to the diagonal part of Eq. \jna{raspisano}:
\vspace{0ex}
\ofa 
2\delta B_1 &=&   i {\epsilon_3}B_1 
 , \cr 
2\delta B_2 &=& - i {\epsilon_3}B_2 
 , \cr 
2\delta B_3 &=& - i {\epsilon_3}B_3 
 , \cr
2\delta B_4 &=&   i {\epsilon_3}B_4
\ . 
\zffa{didiew}

\ni This is a $u(1)$ algebra of the transformations:
\vspace{0ex}
\of 
\delta \left|\psi_{1,2}\right> 
= 
i {\epsilon_3\over 2}{\bf \sigma}_{3} \left|\psi_{1,2}\right> , \zf
\noindent and the corresponding group is $U(1)$. The parameter
space of this group is the circle ${\Bbb S}^1$ of circumference
$4\pi$. That group is the subgroup of both $SU(2)_I$ and $SU(1,1)_I$
groups.

\section{Symmetries of the equations of motion}

\label{section_on_eoms}

\label{subsection_on_eom_esymm_C}

\desnoime{eom\_esymm\_C.tex}
 
In general, one should distinguish the symmetries of the integrals
of motion from the symmetries of the equations of motion.
\vspace{2ex}
\hrule\smallskip

{\defi :
{\bf E-symmetries} \cite{kowalski91}: 
Any
vector-field $\vec{L}$ that satisfies the master equation 
\vspace{0ex}
\of
\left[ {\vec L} , \vec{F} \right]  = 0
 ,
\zff{master2}

\noindent is the symmetry of the dynamical equations \jna{eom}.
}
\smallskip\hrule\medskip

The set ${\it g}_E$ of E-symmetries is also a Lie-algebra, 
i.e. it is linear, 
and the
commutator of any two E-symmetries is another E-symmetry. 
So, one can describe
the full algebra by its generators and their commutation relations.

The E-symmetries are sought in the form of the most general linear
ansatz
\vspace{0ex}
\of
\vec{L} = x^{\mu} a_{\mu}{}^{\nu} \partial_{\nu} .
\zf

\ni After some algebra, six generators are found for the 4WM system:
\vspace{0ex}
\ofa
\vec{L}_0  &=& B_1 \partial_1  + B_2\partial_2 + B_3\partial_3 + B_4\partial_4 + c.c. 
 , \cr
\vec{L}_1  &=& i (B_1\partial_1 + B_2\partial_2 + B_3\partial_3 + B_4\partial_4 - c.c.)
 , \cr
\vec{L}_2  &=& i (B_1\partial_1 - B_2\partial_2 - c.c.)
 , \cr
\vec{L}_3  &=& i (B_3\partial_3 - B_4\partial_4 - c.c.)
 , \cr
\vec{L}_4  &=& \bar{B}_2\partial_1 - \bar{B}_1\partial_2 + \bar{B}_4\partial_3 - \bar{B}_3\partial_4 + c.c.
 , \cr
\vec{L}_5  &=& i (\bar{B}_2\partial_1 - \bar{B}_1\partial_2 + \bar{B}_4\partial_3 - \bar{B}_3\partial_4 -c.c.).
\zffa{esymmetriesgeneral}

\ni These six generators form the complete set of linear E-symmetries
for the general (complex) $\Gamma$. They generate the algebra
${\it g}_E \equiv {\it r}\oplus u(1)\oplus u(1)\oplus su(1,1)$,
with the non-vanishing commutators 
\vspace{0ex}
\ofa
[\vec{L}_1,\vec{L}_4] &=& -2 \vec{L}_5  , \cr
[\vec{L}_1,\vec{L}_5] &=&  2 \vec{L}_4  , \cr
[\vec{L}_4,\vec{L}_5] &=&  2 \vec{L}_1  .
\zfa

Defining the general infinitesimal E-symmetry by $\nedelta$ $\equiv$
$\sum_{i=0}^5 \theta_i \vec{L}_i$, we find the transformation law of the 
fields:
\vspace{0ex}
\ofa
\nedelta B_1
&=&
[\theta_0+i(\theta_1+\theta_2)]B_1+(\theta_4+i\theta_5)\bar{B}_2
 , \cr
\nedelta B_2
&=&
[\theta_0+i(\theta_1-\theta_2)]B_2-(\theta_4+i\theta_5)\bar{B}_1
 , \cr
\nedelta B_3
&=&
[\theta_0+i(\theta_1+\theta_3)]B_3+(\theta_4+i\theta_5)\bar{B}_4
 , \cr
\nedelta B_4
&=&
[\theta_0+i(\theta_1-\theta_3)]B_4-(\theta_4+i\theta_5)\bar{B}_3
 ,\zfa
\noindent or, in a more compact notation:
\vspace{0ex}
\ofa
\nedelta |\psi_1\rangle 
&=& 
\left[ (\theta_0+ i\theta_1){\bf 1} +
i\pmatrix{\theta_2&0\cr 0 &\theta_3}\right] |\psi_1\rangle  + 
\cr
&&
\ \ \ \ 
- (\theta_4+i\theta_5)
|\psi_{\bar{2}}\rangle 
 ,
\cr
\nedelta  |\psi_2\rangle 
&=& 
\left[ (\theta_0+ i\theta_1){\bf 1} -
i\pmatrix{\theta_2&0\cr 0 &\theta_3}\right] |\psi_2\rangle  +  
\cr
&&
\ \ \ \ 
- \sigma (\theta_4+i\theta_5)
|\psi_{\bar{1}}\rangle .
\zfa
\ni The parameters $\theta_0$ and $\theta_4$ correspond to the two
noncompact dimensions of the symmetry group ${\cal G}_E$, i.e.
their values are arbitrary real numbers. This is in 
contrast to the rest of
the parameters, which are periodic. So, the group of E-symmetries
${\cal G}_E \equiv \exp {\it g}_E$ is ${\Bbb R} \otimes U(1)^2
\otimes SU(1,1)$.

It is easy to check [from the defining relation \jna{master2}] that the 
E-symmetries always map the integrals of motion to the integrals of motion 
(and also the solutions of EOM to the solutions of EOM). Hence:
\vspace{0ex}
\ofa
\nedelta q_1 
&=& 
2\theta_0 q_1 + (\theta_4+i\theta_5)\bar{q}_4+
(\theta_4-i\theta_5)q_4
 , \cr
\nedelta q_2 
&=& 
2\theta_0 q_2 - (\theta_4+i\theta_5)\bar{q}_4-
(\theta_4-i\theta_5)q_4
 , \cr
\nedelta q_3 
&=& 
[2\theta_0 +i (\theta_2+\theta_3)]q_3 
 , \cr
\nedelta q_4 
&=& 
2(\theta_0 +i \theta_1) q_4 +
(\theta_4+i\theta_5)(q_2-q_1)
 .
\zfa
\ni This is a version of the N\"other 
theorem: If one of IOM is taken as the "Hamiltonian" $H$ 
of the system, then the action $\vec{L}_i(H)$
of each $\vec{L}_i$ on such a Hamiltonian 
produces another IOM.

Here, the following linear combinations of regular IOM are forming the irreducible 
representations of the $u(1)_{L_0}\oplus su(1,1)_{L_1,L_4,L_5}$ algebra 
under the E-symmetries: $q_1+q_2$, $q_3$ and $\bar{q}_3$ form singlets
\vspace{0ex}
\ofa
\nedelta (q_1+q_2) 
&=& 
2\theta_0 (q_1+q_2),
\cr
\nedelta q_3 
&=& 
[2\theta_0 +i (\theta_2+\theta_3)]q_3,
\cr 
\nedelta \bar{q}_3 
&=& 
[2\theta_0 -i (\theta_2+\theta_3)]\bar{q}_3, 
\zfa

\ni while 
$|T\rangle :\equiv\left\{q_4,({q_1-q_2})/{\sqrt{2}},\bar{q}_4\right\}^T$ 
is transforming as the triplet representation $\nedelta|T\rangle ={\bf P}|T\rangle $,
where ${\bf P}$ is given by
\vspace{0ex}
$$
\pmatrix{
2(\theta_0+i\theta_1) & 
-\sqrt{2}(\theta_4+i\theta_5)& 
0 \cr
\sqrt{2}(\theta_4-i\theta_5)& 
2\theta_0 &
\sqrt{2}(\theta_4+i\theta_5)\cr
0 &
-\sqrt{2}(\theta_4-i\theta_5)&   
2(\theta_0-i\theta_1)
}.
$$
\ni Hence, one can start from the knowledge of only $q_1$ 
and recover the (almost) full set of
regular integrals $\{q_1,q_2,q_4,\bar{q}_4\}$, 
by acting on them with the E-symmetries.

\subsection{$\Gamma\in {\Bbb R}$}

\desnoime{eom\_esymm\_R.tex}
 
In the $\Gamma\in{\Bbb R}$
case, one can apply the same linear ansatz as in the general case. 
The set of linear E-symmetries thus obtained is 
$\{\vec{L}_0, \cdots, \vec{L}_7\}$, where $\vec{L}_{0-5}$ are 
the already known generators \jne{esymmetriesgeneral}, 
and the two additional generators are found for the RG case:
\vspace{0ex}
\ofa
\vec L_6  
&=& 
B_4\partial_1 - B_3\partial_2 + B_2\partial_3 - B_1\partial_4 + c.c.
 , \cr
\vec L_7  
&=& 
i (\bar{B}_4\partial_1 + \bar{B}_3\partial_2 + \bar{B}_2\partial_3 + \bar{B}_1\partial_4 - c.c.).
\cr 
&&
\zfa
\vspace{0ex}
\ni An alternative approach is to perform the 
redefinition $z\rightarrow \theta(z)$ 
of the "time" variable  
(introduced in the previous section), making the  matrix  
${\bf \tilda{m}}=\pmatrix{0&\sigma\nu\cr -\bar{\nu}&0}$ 
constant. This allows one to translate the
vector-field language (applicable in the general case) into the
matrix language. Define the column
\vspace{0ex}
\of
\left|\Psi\right>\equiv\left(\matrix{
\left|\psi_1\right>\cr
\left|\psi_2\right>\cr
\left|\psi_{\bar{1}}\right>\cr
\left|\psi_{\bar{2}}\right>
}\right) .
\zff{velikispinor}

\ni The evolution equation \jna{dirac} 
can now be written as
\vspace{0ex}
\of
\partial_\theta \left|\Psi\right>={\bf M}\left|\Psi\right> ,
\zff{velikidirac}
\noindent where the constant evolution matrix is
\vspace{0ex}
\of
{\bf M} :\equiv 
{\bf 1}_4 \otimes {\bf \tilda{m}} = 
\left(\matrix{
{\bf \tilda{m}}       &0&0&0 \cr
0&  {\bf \tilda{m}}&      0&0  \cr
0&0&  {\bf \tilda{m}}      &0  \cr
0&0&0&  {\bf \tilda{m}}
}\right) .
\zf

The master equation \jna{master2} now has the matrix form
\vspace{0ex}
\of
[{\bf K},{\bf M}]=0 ,
\zf

\noindent where the matrix ${\bf K}=\left({\bf K}_{\mu\nu}\right)$ 
defines the infinitesimal symmetry of the big "spinor" 
$\nedelta |\Psi\rangle =  {\bf K}|\Psi\rangle$ (here $\mu,\nu\in\{1,2,\bar{1},\bar{2}\}$). 
The above master equation is translated  
into "smaller" versions $[{\bf K}_{\mu\nu},{\bf \tilda{m}}]=0$, valid for 
each $2\times 2$ block matrix ${\bf K}_{\mu\nu}$. 
Solutions of these "small" master equations are all of the same form
\vspace{0ex}
\of 
{\bf K}_{\mu\nu} = \alpha_{\mu\nu}{\bf 1}_2+\beta_{\mu\nu} {\bf \tilda{m}}
, \ \ \ \ \    
(\forall \mu,\nu).
\zf

The analiticity conditions $\left|\psi_{\bar{i}}\right>
= {\bf n}_1{\bf n}_2 \left|\bar{\psi}_i\right>$ 
yield the constraints
\begin{eqnarray}
\alpha_{\bar{i}j} = -\sigma \overline{\alpha_{i\bar{j}}},
&&
\beta_{\bar{i}j}  = -\sigma \overline{\beta_{i\bar{j}}},
\cr
\alpha_{\bar{i}\bar{j}} = \overline{\alpha_{ij}},
&&
\beta_{\bar{i}\bar{j}}  = \overline{\beta_{ij}},
\end{eqnarray}
\ni i.e. 
\vspace{0ex}
\ofa
{\bf K}_{\bar{i}j} 
&=& 
-\sigma \overline{\alpha_{i\bar{j}}} {\bf 1}_2
-\sigma \overline{\beta_{i\bar{j}}} {\bf \tilda{m}},\cr
{\bf K}_{\bar{i}\bar{j}} 
&=& 
\overline{\alpha_{\bar{i}\bar{j}}} {\bf 1}_2
+ \overline{\beta_{\bar{i}\bar{j}}} {\bf \tilda{m}}.
\zfa

\ni Hence:
\vspace{0ex}
\ofa
\nedelta \left|\psi_i\right> 
&=&
\alpha_{ij} \left|\psi_j\right> 
+
\beta_{ij} {\bf \tilda m} \left|\psi_j\right> 
+
\cr
&&
\ \ \ 
+
\alpha_{i\bar{j}} \left|\psi_{\bar{j}}\right> 
+
\beta_{i\bar{j}} {\bf \tilda m} \left|\psi_{\bar{j}}\right>, 
\cr
\nedelta \left|\psi_{\bar{i}}\right> 
&=&
- \sigma \overline{\alpha_{i\bar{j}}} \left|\psi_j\right> 
- \sigma \overline{\beta_{i\bar{j}}} {\bf \tilda m} \left|\psi_j\right> 
+
\cr
&& 
\ \ \ 
+
\overline{\alpha_{ij}} \left|\psi_{\bar{j}}\right> 
+
\overline{\beta_{ij}} {\bf \tilda m} \left|\psi_{\bar{j}}\right>.
\zfa

\ni In this way the rescaled EOM have $32$ symmetries, characterized by the
the real and the imaginary parts of the parameters 
$\{\alpha_{ij},\beta_{ij},\alpha_{i\bar{j}},\beta_{i\bar{j}}\}$ ($i,j=1,2$).

\subsection{On the relation between I-symmetries and E-symmetries}

\desnoime{eom\_esymm\_vs\_isymm.tex}
 
One may ask the question, what is the relation between the two groups of symmetries: 
I-symmetries and E-symmetries? The following general consideration 
clarifies this issue a bit. Let $\vec{F}$ be the EOM vector field, 
$\delta$ an arbitrary E-symmetry,  
$\nedelta$ an arbitrary I-symmetry, and $q$ an arbitrary IOM. 
Since $[\vec{F},\nedelta]\equiv 0$ and $\vec{F}(q)\equiv 0$, 
it follows that $\vec{F}(\nedelta q)=0$, i.e. $\nedelta q \sim q$
(N\"other theorem: E-symmetry of IOM is also IOM). 
From this conclusion and from 
$\delta q\equiv 0$ it follows that $[\nedelta,\delta]q = 0$, i.e. 
$[\nedelta,\delta]\sim\delta$ (E-symmetry maps an I-symmetry into 
another I-symmetry).
Hence, one expects that $[\vec{L}_i,\vec{S}_a]\sim\vec{S}_b$.

This can be explicitly checked in the case of 4WM system: 
the generators $\vec{L}_i$ for $i\in\{0,1,4,5\}$
commute with all three $\vec{S}_a$ generators, whereas 
the remaining two E-symmetries $\vec{L}_{2,3}$ have nontrivial commutators 
with $\vec{S}_a$:
\vspace{0ex}
\ofa
[\vec{L}_2,\vec{S}_1\pm\vec{S}_2] 
&=& 
\mp i \left(\vec{S}_1\pm\vec{S}_2\right) 
 , \cr
[\vec{L}_3,\vec{S}_1\pm\vec{S}_2] 
&=& 
\pm i \left(\vec{S}_1\pm\vec{S}_2\right) 
 , \cr
[\vec{L}_{2,3},\vec{S}_3] 
&=& 
0.
\zfa 
\ni Thus the $su(1,1)_E$ symmetry (generated by $\{\vec{L}_1,\vec{L}_4,\vec{L}_5\}$) 
commutes with the $su(2)_I/su(1,1)_I$ symmetry.

\section{Solution procedures}

\label{section_on_sols}

\subsection{$\Gamma\in{\Bbb R}$: The first procedure}

\desnoime{sol\_R\_1.tex}

We present in detail the solution procedures for the case when $\Gamma$ is real.
This case is physically the most relevant. The case when $\Gamma$ is imaginary, is treated similarly.

The equations for $\alpha_1$  and $\alpha_2$, extracted from
Eqs. \jna{eom}, form  a  closed system of equations:
\vspace{0ex}
\ofa 
2 I \alpha'_1  
&=& 
- \Gamma \left[ q_1 s( \sigma,2\alpha_1 ) 
+ q_2 s( \sigma,2\alpha_2 ) \cos(\Phi) \right]
 ,\cr
2 I \alpha'_2  
&=& 
- \Gamma \left[ q_1 s( \sigma,2\alpha_1 )\cos(\Phi)  
+ q_2 s( \sigma,2\alpha_2 )\right]
 ,
\cr &&
\zffa{samodve}

\noindent which can be integrated with little difficulty. Once
$\alpha_1$ and $\alpha_2$ are known, the remaining four angles are
found easily:
\vspace{0ex}
\ofa 
2 I \beta_1'  &=& 
-\sigma \Gamma q_2 \sin(\Phi) 
s(\sigma,2\alpha_2) t(\sigma,\alpha_1)
 , \cr 
2 I \beta_2' &=& 
-\sigma \Gamma q_1 \sin(\Phi) 
s(\sigma,2\alpha_1) t(\sigma,\alpha_2)
 , \cr 
2 I \gamma_1'  &=& 
-\Gamma q_2 \sin(\Phi) 
s(\sigma,2\alpha_2) ct(\sigma,\alpha_1)
 , \cr 
2 I \gamma_2'  &=& 
-\Gamma q_1 \sin(\Phi) 
s(\sigma,2\alpha_1) ct(\sigma,\alpha_2)
 , 
\zfa

\noindent where $t$ and $ct$ are the remaining two
trigonometric/hyperbolic functions, formed by using the rule
\jna{fje} (see Appendix A). 

Equations \jna{samodve}
are integrated as follows. First, two new variables are
introduced:
\vspace{0ex}
\of 
x_1  = c(\sigma,2\alpha_1) 
 , \ \ 
x_2  = c(\sigma,2\alpha_2)
 . 
\zf

\ni In terms of these variables $q_5 = q_1 x_1 + q_2 x_2$ , and
Eqs. \jna{samodve} become
\vspace{0ex}
\ofa 
I x'_1  &=& 
\Gamma \left[ q_1  + q_2 c(\sigma,\rho)- x_1 q_5 \right]
 , \cr 
I x'_2  &=& 
\Gamma \left[ q_2  + q_1 c(\sigma,\rho) - x_2 q_5 \right]
 . 
\zffa{opetdve}

\ni Note that, due to symmetry, only one of  Eqs. \jna{opetdve} is
independent.  The solution of the other is obtained from the
solution of the  first  one by interchanging $q_1$  and $q_2$ .
This, however, holds only when $\Gamma$ is real. On  the other
hand, using Eq. \jna{ku}, Eq. \jna{opa2} can be written as:
\vspace{0ex}
\of
I q_5'  = \Gamma \left(q - q_5^2 \right)  .
\zff{q5eq}

\ni The integration of this equation depends on the geometry. For TG,
the total intensity is constant ($I=q_1 +q_2 $), so that
\vspace{0ex}
\of
\int{dq_5\over q-q_5^2}={\Gamma z\over I}  .
\zf

\ni For RG, the intensity is not constant, and
\vspace{0ex}
\of
\int{q_5 dq_5\over q-q_5^2}={\Gamma z}.
\zff{qpet}

\ni The value of the integral in TG depends on  whether  $q$  is  larger
or smaller than $q_5^2$ . For the first case:
\vspace{0ex}
\of
q_5(z) = \sqrt{q}\; \th \left[ \th^{-1}\left({q_5(0)\over\sqrt{q}}\right)+
{\sqrt{q}\over I} \Gamma z \right]
\zf

\noindent whereas for the second:
\vspace{0ex}
\ofa
q_5(z) 
&=& 
\sqrt{q}\; \cth \left[ \cth^{-1}\left({q_5(0)\over\sqrt{q}}\right)+
{\sqrt{q}\over I} \Gamma z \right].
\cr
&&
\zfa

\ni In RG:
\vspace{0ex}
\ofa
q_5^2(z) 
&=& 
q_5^2(0) \exp(-2\Gamma z) + q \left[1-\exp(-2\Gamma z)\right] .
\cr
&&
\zfa
\ni Once $q_5$  is determined, Eqs. \jna{opetdve} for $x_1$  and $x_2$  (i.e.
$\alpha_1$ and $ \alpha_2$) can be integrated. The problem,
therefore, can be reduced to the determination of one variable. Other
variables can be solved in quadratures. To complete the solution, it
remains to fit boundary conditions. This problem, however, is more
conveniently addressed by an alternative solution procedure.

\subsection{$\Gamma\in{\Bbb R}$: The second procedure}

\desnoime{sol\_R\_2.tex}

Another convenient method for solution of 4WM equations is based
on the linearization procedure (the replacement of the "time"
variable $z$ by the variable $\theta(z) =
\int_0^z|\mu(z')|dz'+\theta_0 $). Then \jna{dirac} remains the
same, but the matrix ${\bf m}\rightarrow 
{\bf \tilda{m}}$ becomes constant
[$\mu\rightarrow \nu=\exp({i\phi})$].
The explicit solution of Eqs.\jna{dirac} is now
\vspace{0ex}
\ofa
\left|\psi_j(\theta) \right> &=& \left(
\matrix{c(\sigma,\Theta)   &  \sigma \nu s(\sigma,\Theta)\cr
      - \bar{\nu} s(\sigma,\Theta) & c(\sigma,\Theta)}
\right)
\left|\psi_{j}(\theta_0)\right> , \cr
&&
\zffa{reseno}
\noindent where $\Theta=\theta-\theta_0$ and $\theta_0$ is
to be determined from the boundary  conditions. The subscript $0$
stands for the quantities evaluated at $z=0$. The matrix in
Eq. \jna{reseno} explicitly displays the $SU$ nature of the
symmetry of solutions, and allows for an easy identification of Euler
angles for the problem: $\alpha=\phi$, $\beta=2\theta$,
$\gamma=-\phi$.  In  this formulation (real $\Gamma$) only one
independent variable ($\theta$) is found necessary. The angle
$\phi$ is fixed by the boundary conditions.

The evaluation of $\theta_0$ is facilitated
by writing $|Q|$ and $q_5$ in terms of $\theta$:
\vspace{0ex}
\of
|Q| = {\sqrt{q}} s(\sigma,2\theta)/2 
 , \ \  
q_5  = -\sigma\sqrt{q} c(\sigma,2\theta) .
\zf

\noindent The form of the solution is different in the two
geometries,  since $I$  is constant in TG, whereas it is not in
RG. At this point the symmetry in treating the two geometries
is broken. The solution of Eq. (\ref{q5eq}) is
\vspace{0ex}
\of
\tan(\theta) = \tan(\theta_0)  
\exp \left({\sqrt{q}}\Gamma z/I\right)  , 
\zff{teta}

\noindent in TG, and
\vspace{0ex}
\of
\sh(2\theta) = \sh(2\theta_0 ) \exp(\Gamma z)  ,
\zf

\noindent in RG. The procedure for evaluation of $\theta_0$
is also different in the two geometries. We first present the TG case.

The angle $\theta_0$  is found when boundary conditions  are  applied  to
the  solution given by Eq. (\ref{teta}). The conditions are that the fields  are
given  at  the opposite faces of the crystal: $A_j (z=0 \ \hbox{ or } \ z=d) =
C_j$. In OPC $C_3 = 0$. Using these conditions, a number of auxiliary
quantities is defined:
\vspace{0ex}
\ofa
u &=& |C_4|^2 -|C_1|^2 +|C_2|^2  , \cr
v &=& |C_4|^2 -|C_1|^2 -|C_2|^2  ,\cr
p &=& 2 C_1 \bar{C}_4 \exp(-i\phi)  , \cr
\alpha &=& \exp\left(-{\sqrt{q}}\Gamma d/I\right)  ,
\zfa

\noindent (all real), and a shorthand notation is introduced:
\vspace{0ex}
\of
x = \tan(\theta_d - \theta_0 )  , \ \ y = \tan(\theta_d + \theta_0 ) .
\zf

\noindent There exists a rational relation connecting $x$ and $y$:
\vspace{0ex}
\of
y = {u x + p \over v  - p x}  ,  \ \ x = {v y - p \over u + p y} .
\zf

\noindent It is seen that $x$ and $y$ depend on $\theta_0$ and
[through Eq. (\ref{teta})] on $q$. However, there also exists a relation
expressing $x$ (and likewise $y$) only through $q$:
\vspace{0ex}
\of 
x = - {c\over q-a } \pm \left[\left({c\over
q-a}\right)^2-{q-b\over q-a}\right]^{1/2} , 
\zf
\noindent where $a= p^2 + u^2$ , $b = p^2 + v^2$ ,
$c=p(u-v)=2p|C_2|^2$ . This relation is used  to  write an
implicit equation for $q$:
\vspace{0ex}
\of 
\alpha = \xi^2  - \eta^2 , 
\zff{alpha}

\noindent where $\xi=(1-\alpha)/2x$, $\eta=(1+\alpha)/2y$. Thus,
given the boundary conditions, Eq. (\ref{alpha}) is to be solved numerically, 
to determine $q$. Given $q$, $x$ and $y$ are found, and $\theta_0$
evaluated from the relation
\vspace{0ex}
\of 
\tan(\theta_0)  = {\alpha\over\xi+\eta} . 
\zf

\noindent This completes the TG procedure.


For RG, one finds two expressions for the modulus of the
grating $|Q|$ at $z=0$:
\vspace{0ex}
\ofa
|Q_0| &=& \th(\Theta_d) {|C|^2/ (e-1)}  = \cr
 &=& |p| \sech\left( |C_1|^2 + |C_4|^2 \right)
 ,
\zffa{q0}

\noindent where $|C|^2 =\sum|C_i|^2$ , $e=\exp(\Gamma d)$, and now
$|p|=|\bar{C}_2 C_4|$. This yields  an expression for
$\sh(\Theta_d)$:
\vspace{0ex}
\of 
\sh(\Theta_d) = {|p|(e-1)\over e\left( |C_1|^2+|C_4|^2 \right)
+|C_2|^2  } . 
\zff{shtheta}

\noindent Using Eqs. (\ref{q0}) and (\ref{shtheta}), an expression for 
$\th(2\theta_0)$ is obtained:
\vspace{0ex}
\of 
\th(2\theta_0) = {\sh(2\Theta_d)\over e-\ch(2\Theta_d)} . 
\zf

\noindent This completes the RG procedure.

\subsection{ $\Gamma\in i{\Bbb R}$}

\desnoime{sol\_iR.tex}

It is useful to note that beside the equations \jna{opa1} and \jna{vazna}, 
one can derive an equation for the intensity
\vspace{0ex}
\of 
I I'  = 2 (\sigma-1) \Re(\Gamma)|Q|^2. 
\zff{opa3}

\ni From these equations it follows that in the $\Gamma$ imaginary case a number of additional 
quantities is constant: $|Q|$, $q_5$, $I$. The equation for the phase can be 
recast as
\vspace{0ex}
\of
I \partial_z\arg (Q) = - \tilda{\Gamma} q_5,
\zf 

\ni where $\Gamma \equiv: i\tilda{\Gamma}$,  and solved: 
\vspace{0ex}
\of
\arg(Q(z)) =\arg (Q(0))- \tilda{\Gamma} \frac{q_5}{I} z. 
\zf

\ni Since $|\mu(z)|={|\Gamma Q(z)|}/{I(z)}={|\tilda{\Gamma}| |Q|}/{I}$ 
is constant, one obtains an explicit expression
\vspace{0ex}
\of
\mu(z) =
|\mu_0| \exp({i\phi_0-i \Omega z}), 
\zf

\ni where 
$|\mu_0|:\equiv {|\tilda{\Gamma}| |Q|}/{I}$,
$\phi_0:\equiv \pi/2 +\arg {\tilda{\Gamma}} +\arg (Q(0))$ and
$\Omega :\equiv {\tilda{\Gamma} q_5}/{I}$.

Now the evolution matrix ${\bf m}(z)$ from the 
spinor EOM \jna{dirac} has the form
\vspace{0ex}
\of
{\bf m}(z) = \pmatrix{
0& \sigma |\mu_0| e^{i\phi_0-i\Omega z}\cr
- |\mu_0| e^{-i\phi_0+i\Omega z} & 0
},
\zf 

\ni and the formal solutions of EOM \jna{dirac} are
\vspace{0ex}
\of
|\psi_i(z)\rangle  = {\bf U}(z) |\psi_i(0)\rangle .
\zff{formalnoresenjehaha}

\ni The ${\bf U}(z)$ matrix is the {\it ordered exponential} 
(see Appendix \ref{section_on_OE_C} for a discussion) of ${\bf m}(z)$:
\vspace{0ex}
\of
{\bf U}(z) = \left(\exp\left({\int_0^z dz' {\bf m}(z')}\right)\right)_+
\zf

\ni where the {\it plus} subscript indicates the path-ordered 
nature of the exponential. In practice, to obtain the explicit 
form of ${\bf U}(z)$ in terms of non-ordered quantities, 
one has to solve the initial value problem (IVP) 
\vspace{0ex}
\ofa
\partial_z {\bf U}(z) &=& {\bf m}(z) {\bf U}(z), \cr
{\bf U}(0) &=& {\bf 1}.
\zfa

For the specific ${\bf m}(z)$ the explicit solution to this IVP 
is found (see Appendix \ref{section_on_solving_iR_OE}):
\vspace{0ex}
\ofa
U_{11} 
&=&  
\exp\left({-i\frac{\Omega z}2}\right)
\left[
\cos\left(\frac{\Xi z}2\right)
+ 
i \frac{\Omega}{\Xi}
\sin\left(\frac{\Xi z}2\right)
\right], 
\cr
U_{12} 
&=&  
i\frac{2\sigma Q}{\sqrt{q}} 
\exp\left({-i\frac{\Omega z}2}\right)
\sin\left(\frac{\Xi z}2\right),
\cr
U_{21} 
&=&  
-i\frac{2\sigma \bar{Q}}{\sqrt{q}} 
\exp\left({i\frac{\Omega z}2}\right)
\sin\left(\frac{\Xi z}2\right),
\cr
U_{22} 
&=&  
\exp\left({i\frac{\Omega z}2}\right)
\left[
\cos\left(\frac{\Xi z}2\right)
- 
i \frac{\Omega}{\Xi}
\sin\left(\frac{\Xi z}2\right)
\right],
\cr
&&
\zfa
\ni
where $\Xi:\equiv {\tilda{\Gamma}\sqrt{q}}/{I}$ and 
$q:\equiv q_5^2+4\sigma|Q|^2$. 
It is easy to check that $\det{\bf U}(z)=1$.

In this manner, for known initial values $|\psi_i(0)\rangle$, the full solution 
at later "times" $z>0$ is given by Eq. \jna{formalnoresenjehaha}. 
However, by the nature of the 4WM system, one knows only the part of initial conditions. The system represents a split boundary value problem.

$\bullet$
For the TG case the beams $B_1=A_1$ and $B_3=A_4$ are entering the crystal sample from the 
$z=0$ side, while other two beams $B_2=A_2$ and $B_4=A_3$ are coming from the 
$z=d$ side. Thus, only $|\psi_1\rangle $ is determined at $z=0$, while $|\psi_2\rangle $ 
has a fixed value at $z=d$ boundary:
\vspace{0ex}
\of
|\psi_1(0)\rangle  = \pmatrix{C_1\cr C_3}, \ 
|\psi_2(d)\rangle  = \pmatrix{C_4\cr -\sigma C_2}.
\zf

\ni In this way:
\vspace{0ex}
\ofa
|\psi_1(z)\rangle  &=& {\bf U}(z) |\psi_1(0)\rangle ,\cr
|\psi_2(z)\rangle  &=& {\bf U}(z) {\bf U}(d)^\dagger |\psi_2(d)\rangle .
\zfa

$\bullet$
For the RG case
only  $B_1=A_1$ and $B_4=A_4$ are known at the 
$z=0$ boundary: $B_1(0)=C_1$, $B_4(0)=C_4$, while the remaining two 
field variables are given at the $z=d$ boundary: $B_2(d)=C_2$, $B_3(d)=C_3$.
This means that both spinors $|\psi_i\rangle $ ($i=1,2$) are satisfying the
mixed boundary conditions, where one component satisfies the 
initial value condition (at $z=0$) and the other component
satisfies the final boundary condition (at $z=d$):
\vspace{0ex}
\ofa
|\psi_1(0)\rangle  = \pmatrix{C_1\cr B_3(0)}, 
&&
|\psi_1(d)\rangle  = \pmatrix{B_1(d)\cr C_3}, 
\cr
|\psi_2(0)\rangle  = \pmatrix{C_4\cr -\sigma B_2(0)}, 
&&
|\psi_2(d)\rangle  = \pmatrix{B_4(d)\cr -\sigma C_2}, 
\cr
&&
\zfa

\ni where $B_3(0)$, $B_1(d)$, $B_2(0)$ and $B_4(d)$ are unknown.
In order to determine them, one has to use the 
evolution formula Eq. \jna{formalnoresenjehaha}
to express the unknown boundary values in terms of the known ones. 
After some simple algebra one obtains
\vspace{0ex}
\ofa
B_1(d) &=& \frac{C_1+U_{12}(d)C_3}{U_{22}(d)}, \cr
B_3(0) &=& \frac{C_3-U_{21}(d)C_1}{U_{22}(d)}, \cr 
B_4(d) &=& \frac{C_4-\sigma U_{12}(d)C_2}{U_{22}(d)}, \cr
B_2(0) &=& \frac{C_2+\sigma U_{21}(d)C_4}{U_{22}(d)}. 
\zfa

\ni Here the unimodality condition $\det{\bf U}(z)=1$ ($\forall z$)
was used. This concludes the solution procedure for both geometries.

\section{ The (Pseudo)Hamiltonian structure of 4WM }

\label{section_on_pseudo}

\desnoime{pseudo.tex}
 
Considering again the  form of the spinorial EOM 
\jna{dirac}, one may notice that (in the TG with $\mu\in{\Bbb R}$ case) 
the matrix ${\bf m}$ is antisymmetric,
resembling the symplectic matrix used in the Hamiltonian formalism 
for mechanical systems. This notice gives rise to the question 
whether it is posible to reformulate the 4WM system as 
a Hamiltonian system. In this section one possible approach to the 
problem is considered. First the neccessary general definitions are given,
and then the specifics of the 4WM system are discussed.
 
\subsection{ Preliminaries}

\desnoime{pseudo\_intro.tex}
 
For the formulation of the Hamiltonian formalism \cite{arnold88} 
one needs a phase space in
the form of a smooth manifold $V$ and a closed nondegenerate
differential 2-form
$\hat{F}$ (the {\it field-strength} form)
defined on it, which endows a 
symplectic structure on $V$. 
In the phase space with the canonical coordinates 
$\{q^1,...q^D,p_1,...p_D\}$ ($q^i$ not to be
confused with the conserved quantities), the canonical form
of $\hat{F}$ is
\vspace{0ex}
\of 
\hat{F}=\sum_{k=1}^D dq_k\wedge dp_k  .
\zf

The 2-form $\hat{F}$ sets up an isomorphism between the tangent 
space $TV$
and the cotangent space ${}^*TV$. Denote the inverse mapping by 
$\hat{J}:{}^*TV\rightarrow TV$. In the canonical coordinates 
$\hat{J}$ has the form
\vspace{0ex}
\of 
\hat{J}=\sum_{k=1}^D \partial_{q^k}\wedge \partial_{p_k} .
\zf

In a general system of coordinates $\{x^\mu|\mu=\overline{1,2D}\}$, 
the forms of $\hat{F}$ and $\hat{J}$ become
\vspace{0ex}
\ofa 
\hat{F}&=&\frac12  F_{\mu\nu} dx^\mu\wedge dx^\nu ,
\cr
\hat{J}&=&\frac12  J^{\mu\nu}\partial_\mu\wedge \partial_\nu ,
\zfa

\noindent with mutually inverse skew-symmetric matrices 
${\bf F}=\left({F}_{\mu\nu}\right)$ and 
${\bf J}=\left(J^{\mu\nu}\right)$. Here the summation over the 
repeated greek indices is assumed. 
The matrix ${\bf J}$ is known as the {\it symplectic}
matrix.

Physical quantities are smooth functions on $V$, forming the space 
$C^\infty(V)$. A Poisson bracket is defined in $C^\infty(V)$,
generating a Lie-algebra structure
\vspace{0ex}
\of 
\{f,g\}_{PB}= {J}^{\mu\nu}\partial_\mu f\partial_\nu g .
\zf

In the canonical coordinates this means
\vspace{0ex}
\of 
\{f,g\}_{PB}=\sum_{k=1}^D 
(\partial_{q^k}f\partial_{p_k}g\ -
\partial_{q^k}g\partial_{p_k}f) .
\zf
The Poisson bracket is bilinear, skew-symmetric, 
and obeys the Jacobi identity,
which is equivalent to the closeness of the 2-form $\hat{F}$: 
$d\hat{F}=0$. 
Later more will be elaborated on this condition.

The dynamics is determined by the choice of the Hamilton 
function $H$ on  $V$. The external differential $dH$ is 
a covector field (1-form), and $\hat{J}\cdot dH$ 
is the corresponding Hamilton's vector field on $V$.
The Hamilton equation of motion is specified by equating the tangent vector field 
$\dot{\vec{x}}:\equiv \dot{x}^\mu \partial_\mu$ with the Hamilton's vector 
field:
\vspace{0ex}
\of 
\dot{\vec{x}} =\hat{J}\cdot dH 
.
\zff{ham}

The Poisson bracket $\{f,g\}_{PB}$ 
may now be represented by the action of the covector
$df$ on Hamilton's vector field $\hat{J}\cdot dg$:
 $\{f,g\}_{PB}=df(\hat{J}\cdot dg)$. Therefore, 
the derivative of function $f$
in the direction of Hamilton's vector field $\hat{J}\cdot dH$ 
is in fact $\{F,H\}_{PB}$. 
Hence, the Hamilton equation (\ref{ham}) can be written as 
$\dot f=\{f,H\}_{PB}$ for an arbitrary function $f$.
Since the coordinate functions $\{q^1,...q^D,p_1,...p_D\}$ 
form a complete basis, the equations
\vspace{0ex}
\ofa 
\dot q^k &=& \{q^k,H\}_{PB}=\partial_{p_k}H,
\cr  
\dot p_k &=& \{p_k,H\}_{PB} = -\partial_{q^k}H, 
\zfa

\noindent form a closed system. 
These are the canonical Hamilton equations of motion.

\subsection{ Four-wave mixing}

\desnoime{pseudo\_4wm\_C}
 
In an attempt to cast the 4WM system in the Hamiltonian form, one
encounters several problems. 

First, there is no clear choice of the Hamiltonian 
(Hamilton's function) $H(x)$. 
It can be an arbitrary real function of the full set of conserved 
quantities: $H(x)=h(q(x))$. 
For example, for the general $\Gamma\in{\Bbb C}$,
one can identify three convenient families of Hamiltonians 
\vspace{0ex}
\ofa 
H_{Q1, \epsilon} &=& q_1+ \epsilon q_2  , \cr 
H_{Q3 \lambda}  &=& \lambda q_3 + \bar{\lambda} \bar{q}_3 , \cr 
H_{Q4 \lambda}  &=& \lambda q_4 +  \bar{\lambda} \bar{q}_4  ,  
\zfa
\ni where $\epsilon=\pm 1$ and  $|\lambda|=1$. Clearly, these
families are not exhausting all the possible choices, even among the 
Hamiltonians that are linear in the regular IOM. 
 
The corresponding symplectic matrices are
\vspace{0ex}
{\small
\ofa
{\bf J}_{q 1 \epsilon} 
&=& 
\pmatrix{ 
&&&& &&\mu&\cr
&&&& &&& \epsilon           \bar{\mu}\cr
&&&& -\bar{\mu}&& &\cr
&&&& &- \epsilon           \mu & &\cr
&&\bar{\mu}&  &&&&\cr
&&& \epsilon           \mu &&&&\cr
-\mu& & &   &&&&\cr
&- \epsilon           \bar{\mu}& &     &&&&}
 ,
\cr
&&
\zffa{icybicyspider}

$$
{\bf J}_{q 3 \lambda} = 
\pmatrix{ 
&&&&&-\sigma\bar{\lambda}\mu&& \cr
&&&&\sigma\lambda\bar{\mu}&&&  \cr
&&&&&&&-\bar{\lambda}\bar{\mu}   \cr
&&&&&&\lambda\mu&      \cr
&-\sigma\lambda\bar{\mu}&&&&&& \cr
\sigma\bar{\lambda}\mu&&&&&&&  \cr
&&&-\lambda\mu&&&&     \cr
&&\bar{\lambda}\bar{\mu}&&&&&}
 ,
$$

$$
{\bf J}_{q 4 \lambda} = 
\pmatrix{ 
&&&\bar{\lambda}\mu    & &&& \cr
&&\bar{\lambda}\bar{\mu}&   & &&& \cr
&-\bar{\lambda}\bar{\mu}&&  & &&& \cr
-\bar{\lambda}\mu& &&   & &&& \cr
&&&& &&& \lambda \bar{\mu}     \cr
&&&& && \lambda \mu&    \cr
&&&& &- \lambda \mu&&   \cr
&&&& - \lambda \bar{\mu}& &&    }
 .
$$

}

\noindent 
The matrix elements that are not written explicitly, are zero.
Note that for these three families (subscript $K$ is the index of 
each individual family):
\vspace{0ex}
\of
{\bf J}_{K}^{-1} = {1\over |\mu|^2} {\bf J}_{K}^{\dag} ,\qquad
{\bf J}_{K}^T=-{\bf J}_{K}.
\zf

\ni These matrices are chosen in such a way to satisfy the three basic
requirements: they are antisymmetric, nonsingular (in the matrix sense), 
and they give the same EOM
$$ J_K^{\mu\nu} \partial_\nu H_K = f^\mu(x),$$
\ni where $f^\mu(x)$ is the right-hand side of the EOM \jna{eom}:
$\dot{x}^\mu=f^\mu$. The fact that all Hamiltonian structures 
must reproduce the same dynamical equations
\jna{eom} means that for any two structures $(H_{A},{\bf J}_{A})$
and $(H_{B},{\bf J}_{B})$ one can write:
\vspace{0ex}
\of
\babla{BA} H_{A}=\nabla H_{B} ,
\zff{rieman}
\noindent where $\babla{BA}:\equiv {\bf R}_{BA}\cdot\nabla$ 
defines {\it the cogradient} and 
${\bf R}_{BA} :\equiv {\bf J}_{B}^{-1}{\bf J}_{A}$ is
{\it the recursion matrix}, 
connecting the Hamiltonian structures $(A)$ and $(B)$.

Equation \jna{rieman} defines the mapping from the first Hamiltonian
structure $(A)$ to the second one $(B)$. It is interesting to assume 
for a moment that there exists some function $H_C$ 
whose gradient is the $BA$-cogradient
of $H_{B}$. If such a function exists, it will be an integral
of motion:
\begin{eqnarray}
\partial_z H_C 
&=&
\dot{x}^\mu \partial_\mu H_C 
=
\cr
&=&
J_B^{\mu\nu}\partial_\nu H_B (J_B^{-1})_{\mu\alpha}
J_A^{\alpha\beta}\partial_\beta H_B
=
\cr
&=&
-\partial_\alpha H_B J_A^{\alpha\beta} \partial_\beta H_B \equiv 0.
\nonumber
\end{eqnarray}
\noindent Thus, the conserved quantity $H_C$, if it exists,
is a new Hamiltonian of the system, and the corresponding 
symplectic matrix is ${\bf J}_C :\equiv {\bf R}_{AB}{\bf J}_B
= {\bf J}_B{\bf J}_A^{-1}{\bf J}_B$. 

One can continue along the same lines, defining a series of conserved
quantities ("Hamiltonians") and the corresponding symplectic matrices:
\vspace{0ex}
\of
\matrix{
H_{A} &\rightarrow &
H_{B} &\rightarrow & 
H_{C} &\rightarrow & 
\cdots \cr
\hat{J}_{A} &\rightarrow &
\hat{J}_{B} &\rightarrow &
\hat{J}_{C} &\rightarrow & 
\cdots
}
\zf

\ni The sequence terminates 
when the Hamiltonians start repeating themselves (i.e. they became linear combinations of the 
previous members of the sequence). 
This type of sequence of the Hamiltonians and the corresponding symplectic matrices
is common in the two-dimensional integrable systems. There exists a 
{\it multi-Hamiltonian property} of integrable systems, 
whereby the chain of Hamiltonians 
is (usually) non-terminating, leading to an infinite set of non-equivalent 
IOM, and, thus, to the complete integrability of the system. 

In 4WM one expects that all such sequences, if they exist at all, 
should terminate after a few terms. For example,  consider the two 
structures $(H_A:\equiv H_{Q1,+1}=q_1+q_2,J_A)$ and 
$(H_B:\equiv H_{Q1,-1}=q_1-q_2,J_B)$, where $J_A$ and $J_B$ 
are the special cases of \jna{icybicyspider}. 
The recursive matrix is 
${\bf R}_{BA}:\equiv {\bf J}_B^{-1}{\bf J}_A 
= \hbox{diag}(1,-1,1,-1,1,-1,1,-1)$ and the basic identity is 
${}^\dagger\nabla_{BA} H_A = \nabla H_B$. The definition of the induced 
$H_C$ is
\vspace{0ex}
\of
\nabla H_C :\equiv {}^\dagger\nabla_{BA} H_B,
\zf 
\ni and its solution is $H_C=q_1+q_2$. Thus $H_C=H_A$ and the
sequence is periodic: 
$H_A \rightarrow H_B \rightarrow H_A \rightarrow \cdots$.

Each Hamiltonian structure $(H_{K},{\bf J}_{K})$ has 
the corresponding Poisson bracket:
\vspace{0ex}
\of
\{f,g\}_{K} :\equiv (\nabla f)^T{\bf J}_{K}(\nabla g) .
\zf
\ni This bracket is  antisymmetric 
$\{g,f\}_K = -\{f,g\}_K$, 
and can be characterized by the set of 
basic brackets:
\vspace{0ex}
\of
\{x^\mu,x^\nu\}_{K} = J_{K}^{\mu\nu}.
\zf 
\ni However, a bracket so defined does not
satisfy the expected Jacobi identity 
$\{f,\{g,h\}_K\}_K + \hbox{cyclic} \equiv 0$.
The tensor of {\it non-Jacobianity} measures the 
violation of the Jacobi identity:
\vspace{0ex}
\ofa 
\Omega_{(K)}^{\mu\nu\alpha} 
&:\equiv& 
\{\{x^{\mu},x^{\nu}\}_K,x^{\alpha}\}_K + {cyclic}
=
\cr
&=&
(\partial_\sigma J_{K}^{\mu\nu}) J^{\sigma\alpha} + {cyclic}.
\zfa
\noindent This tensor 
is essentially the same as the 
tensor of deflection from the 
Bianchi identity, 
${\omega}:\equiv d\hat{F}$ 
(i.e. 
$\omega_{\mu\nu\alpha}:\equiv \partial_\mu F_{\nu\alpha} + cyclic$):
\vspace{0ex}
\of 
\Omega_{(K)}^{\mu\nu\alpha} 
= J_{K}^{\mu\mu'} J_{K}^{\nu\nu'} J_{K}^{\alpha\alpha'}
\omega_{(K)\mu'\nu'\alpha'}
 . 
\zf

\noindent In the case of non-singular ${\bf J}$,
the Jacobi condition is equivalent to the
Bianchi identity.

In the system at hand,  
the fact that $\Omega_{(K)}$ is not 
disappearing
is the consequence of the non-constancy of $\mu$. 
For example, take again the Hamiltonian $H=H_{Q1,+1}=q_1+q_2$. Then
\vspace{0ex}
\of
{\bf J}=  \mu {\bf E}_+ + \bar{\mu} {\bf E}_- ,
\zf
\ni where 

$$
{\bf E}_+= \pmatrix{
&&&&&&1&\cr
&&&&&&&0\cr
&&&&0&&&\cr
&&&&&-1&&\cr
&&0&&&&&\cr
&&&1&&&&\cr
-1&&&&&&&\cr
&0&&&&&&
},
$$

$$
{\bf E}_-= \pmatrix{
&&&&&&0&\cr
&&&&&&&1\cr
&&&&-1&&\cr
&&&&&0&&\cr
&&1&&&&&\cr
&&&0&&&&\cr
0&&&&&&&\cr
&-1&&&&&&
}.
$$

\ni Note that ${\bf E}_\pm^T=-{\bf E}_\pm$, 
${\bf E}_\pm \cdot {\bf E}_\mp =0$, and 
\vspace{0ex}
\of
\left({\bf E}_\pm\right)^2= 
-\frac12 \left[ {\bf 1}_8 \pm 
{\bf \sigma}_3 \otimes {\bf \sigma}_3 \otimes {\bf \sigma}_3
\right].
\zf

\ni As a simple consequence of these expressions, the following identity
is valid
\vspace{0ex}
\of
\left(
\alpha {\bf E}_+ + \beta {\bf E}_-
\right)^{-1}
=
-\frac1\alpha {\bf E}_+ -\frac1\beta {\bf E}_-,
\zf

\ni for arbitrary (nonzero) $\alpha$ and $\beta$. This identity is 
used to evaluate the "field strength" matrix:
\vspace{0ex}
\of
{\bf F}= -\frac1\mu {\bf E}_+ - \frac1{\bar{\mu}} {\bf E}_- .
\zf

From this, one obtains 
\vspace{0ex}
\ofa
\omega_{\mu\nu\alpha} 
&=&
\frac1{\mu^2}
\left[
(\partial_\alpha \mu) (E_{+})_{\mu\nu} + cyclic
\right]
+ \cr
&&
\ \ \ \
+ \{\mu\rightarrow\bar\mu, E_+\rightarrow E_-\}.
\zfa  

\ni For example, $\omega_{B_1,B_4,\bar{B}_2}=-\frac1{\mu^2}\partial_1\mu$. 
Thus, the Bianchi identity is clearly broken, and one can not find the 
potential ${\cal A}_\mu(x)$ such that $F_{\mu\nu}(x)$ is its strength
tensor.

One can see the dual nature of the same obstacle, expressed in terms of the 
Poisson bracket, in the following way. The basic brackets are
\vspace{0ex}
\ofa
\left\{B_1,\bar{B}_3\right\}_{PB} &=&\mu,
\cr
\left\{B_2,\bar{B}_4\right\}_{PB} &=&\bar{\mu}.
\zfa
\ni and one non-vanishing component of $\Omega$ is
\vspace{0ex}
\of
\Omega^{B_1\bar{B}_3B_2}=-\bar{\mu}\bar{\partial}_4 \mu.
\zf

\ni The full list of non-vanishing components of $\Omega$ is given in 
the Appendix \ref{appendix_with_components_of_Omega}. Thus, the Poisson
bracket is not self-consistent: one can not apply it consecutively
on the phase space without running into inconsistencies.  
This is the second, and much more serious problem with the presented 
approach to casting the 4WM system in the Hamiltonian form.
One can say that the 4WM system has a {\it pseudo-Hamiltonian structure}. 

If all components of $\omega$ were zero, 
one would be able to find the potential
functions ${\cal A}_\mu$ of strength tensor $F_{\mu\nu}=\partial_\mu 
{\cal A}_\nu - \partial_\nu {\cal A}_\mu$. Then the system could be 
formulated as the Lagrangian one, with the action functional 
$S(z_1,z_2):\equiv \int_{z_1}^{z_2}dz L(x,\dot{x})$, where the
Lagrangian function is
\vspace{0ex}
\of
L(x,\dot{x}):\equiv \dot{x}^\mu{\cal A}_\mu(x) -H(x).
\zf

\ni The Euler-Lagrange EOM 
corresponding to this Lagrangian are the Hamilton equations 
\jna{ham}. The elements of the Lagrangian formalism are provided in Appendix
\ref{app_lagrange}. 

Since $\omega\ne 0$,
one may search for the solutions in the form
\vspace{0ex}
\of
F_{\mu\nu} = f (\partial_\mu{\cal A}_\nu - \partial_\nu{\cal A}_\mu),
\zf
\ni which leads to 
\vspace{0ex}
\of
(\partial_\alpha f) F_{\mu\nu} + cyclic= f \omega_{\alpha\beta\gamma}.
\zf

\ni The direct consequence of this equation is 
\vspace{0ex}
\of
\partial_\alpha \ln{f} = \frac1{2D-2} \omega_{\alpha\mu\nu}J^{\nu\mu}
\zf 

\ni where $2D=8$ in the 4WM system. After some algebra one derives
\vspace{0ex}
\ofa
\partial_\alpha \ln{f} =-\partial_\alpha \ln{|\mu|}+\frac{i}3 
\left(
{\bf \sigma}_3\otimes{\bf \sigma}_3\otimes{\bf \sigma}_3
\right)_{\alpha}{}^{\nu} 
\partial_\nu \phi.
\cr
&&
\zffa{smislu}

\ni where $\phi=\arg(\mu)$.

\subsection{The  $\Gamma \in {\Bbb R}$ case}

\desnoime{pseudo\_4wm\_R.tex}
 
If $\phi$ is constant 
(i.e. $\Gamma\in{\Bbb R}$), 
the solution of Eq. \jna{smislu} is $f=1/{|\mu|}$. 
Then 
$$
\tilda{F}_{\mu\nu} =
- \bar{\nu} \left(E_+\right)_{\mu\nu}
- 	 \nu  \left(E_-\right)_{\mu\nu},
$$
\ni where 
$\tilda{F}_{\mu\nu}:\equiv F_{\mu\nu}/f = 
\partial_\mu {\cal A}_\nu - \partial_\nu {\cal A}_\mu$,  
and the solution for the potential ${\cal A}_\mu$ is
\vspace{0ex}
\of
{\cal A}_\mu = 
- \frac12
\left[
\bar{\nu} \left(E_+\right)_{\mu\nu}
+ 	 \nu  \left(E_-\right)_{\mu\nu}
\right] x^\nu
,
\zf   

\ni with $\nu=\exp(i \phi)$ (not to be confused with the index $\nu$).

To construct the action for this case one has to go one step back. The  
factorization of $f=1/{|\mu(x)|}$ from $F_{\mu\nu}$ is equivalent 
to the introduction of a new time parameter $\theta(z):\equiv 
\int_0^z dz' {\cal M}(z') + const.$ into EOM, where 
${\cal M}(z):\equiv |\mu\left(x(z)\right)|$ is the on-shell 
value of $|\mu(x)|$. Thus 
$$
\tilda{F}_{\mu\nu} \frac{dx^\nu}{d\theta} = \partial_\mu H(x).
$$
\ni The constant "field-strength" form $\hat{\tilda{F}}$ is closed, 
and  its tensor of non-Jacobianity $\tilda{\omega}$ disappears.
So, one can construct the action in the rescaled time 
\vspace{0ex}
\of
S(\theta_1,\theta_2) 
= 
\int_{\theta_1}^{\theta_2} d\theta
\left[
\frac12 x^\mu \tilda{F}_{\mu\nu} \frac{dx^\nu}{d\theta} -H(x)
\right].
\zf

\ni Further transformation of this action 
\vspace{0ex}
\ofa
S &=&
\int_{z_1}^{z_2} dz {\cal M}(z) 
\left[
\frac12 x^\mu \tilda{F}_{\mu\nu} 
\frac1{{\cal M}(z)}
\frac{dx^\nu}{dz} -H(x)
\right]=
\cr
&=&
\int_{z_1}^{z_2} dz 
\left[
\frac12 x^\mu \tilda{F}_{\mu\nu} 
\frac{dx^\nu}{dz} - {\cal M}(z) H(x)
\right],
\nonumber
\zfa

\ni leads to the Lagrangian 
\vspace{0ex}
\ofa
L &=& 
- \frac12
\dot{x}^\mu
\left[
\bar{\nu} \left(E_+\right)_{\mu\nu}
+ 	 \nu  \left(E_-\right)_{\mu\nu}
\right] x^\nu
+
\cr
&&
\ \ \ 
- {\cal M}(z) (q_1 + q_2),
\zffa{strange_strange}
\ni  Note that if in the above expression ${\cal M}(z)$ is directly
replaced by $|\mu(x)|$, the obtained corresponding variation equations 
are wrong. 

In the $\Gamma\in{\Bbb R}$ case the set of IOM is enlarged by the 
"exceptional" IOM $\{w_{1-6}\}$ (and their complex conjugates), 
and one can construct some additional families of 
(linear in IOM) Hamiltonians ($ \epsilon=\pm 1$, $|\lambda|=1$,
$|\theta|=1$):
\vspace{0ex}
\ofa
H_{W 1, \epsilon} &=& w_1+ \epsilon  \sigma   w_2  ,\cr
H_{W 3, \lambda}  &=& \lambda w_3 +  \bar{\lambda} \bar{w}_3  , \cr
H_{W 5, \lambda}  &=& \lambda w_5 +  \bar{\lambda} \bar{w}_5  , \cr
H_{W 4, \lambda\theta} &=& \frac12 (\lambda w_4 +  \bar{\lambda} \bar{w}_4  +  \theta     w_6 +  \bar{\theta}
\bar{w}_6)\ .
\zffa{wolfpeter}
\noindent The corresponding symplectic matrices are
{\small
$$
{\bf J}_{W 1 \epsilon           } = 
\pmatrix{
&&&&  \sigma          &&&           \cr
&&&& & \epsilon           \sigma          &&        \cr
&&&& && 1     &         \cr
&&&& &&&  \epsilon          \cr
- \sigma          &&&     & &&&      \cr
 &- \epsilon           \sigma          &&  & &&&      \cr
&& -1 &      & &&&      \cr
&&& - \epsilon      & &&&
},
$$
$$
{\bf J}_{W 3 \lambda             } = 
\pmatrix{
&&&& &&&- \lambda \sigma      \cr
&&&& &&-\bar{\lambda}\sigma&       \cr
&&&& & \lambda \sigma   &&    \cr
&&&& \bar{\lambda} \sigma   &&&    \cr
&&&-\bar{\lambda} \sigma   &&&&    \cr
&&-\lambda\sigma& &    &&&    \cr
&\bar{\lambda}\sigma& &&    &&&    \cr
\lambda\sigma& &&&     &&&
},
$$
$$
{\bf J}_{W 5\lambda} = 
\pmatrix{ 
&-\sigma\bar{\lambda} &&  &&&&  \cr 
\sigma\bar{\lambda}&  &&  &&&&  \cr 
&&&-\bar{\lambda}&         &&&  \cr 
&&\bar{\lambda}&         & &&&  \cr 
&&& &&-\sigma \lambda  &&  \cr 
&&& &\sigma\lambda&    &&  \cr
&&&&& && \lambda           \cr 
&&&&& &-\lambda& } ,
$$

$$
{\bf J}_{W 4\lambda\theta} =
\pmatrix{
&&          \bar{\lambda}&    & &&&        \cr
&&          &\sigma\bar{\theta} & &&&      \cr
-\bar{\lambda}&&      &      & &&&         \cr
&-\sigma\bar{\theta}&   &      & &&&       \cr
&&&  &&& \lambda&                     \cr
&&&  &&& & \sigma\theta               \cr
&&&  &-\lambda&&&                     \cr
&&&  &&-\sigma\theta&&
} .
$$
}
\ni These matrices ${\bf J}_{W \cdots}$ do not depend on 
$\mu$, i.e. they are constant. 
So, their Poisson brackets  satisfy the Jacobi identity. 
This is the case not only for the real, but also for the
complex  coupling $\Gamma$. However, the Hamiltonians \jne{wolfpeter}
are not IOM for the complex $\Gamma$.

To sum up the results, 
for a general $\Gamma$ two types of pseudo-Hamitonian structures
exist:

$\bullet$ \underline{$(H_Q,{\bf J}_Q)$}: 
Hamiltonians $H_Q$ are linear in regular IOM. They are conserved 
quantities in general case, but the corresponding field-strength forms
${\hat{F}_Q}$ are non-closed ${d\hat{F}_Q}\ne  0$. The defect of
this type is the non-closeness of its symplectic structure. 

$\bullet$ \underline{$(H_W,{\bf J}_W)$}:
Hamiltonians $H_W$ are linear in exceptional 
($\Gamma\in{\Bbb R}$) IOM. The corresponding field-strength forms
${\hat{F}_W}$ are constant and closed  in general case. The defect of this 
type of structure is the nonconservation of $W$-Hamiltonians (in general, 
$\Gamma\in{\Bbb C}$ case).

In the case of real coupling, both defects disappear, 
the first one after
rescaling $z\rightarrow\theta(z)$, and the second one because
$W$-Hamiltonians become constant. Then one can construct a consistent
Hamiltonian structure for the 4WM system.


\section{Conclusions}

\label{section_with_conclusions}

\desnoime{conclusions.tex}
 
In summary, the algebraic structures of the 4WM equations in PR crystals
were studied.

First, the form of the equations of motion was used to 
group the basic fields into two doublets, leading to the new 
form of EOM, resembling the Dirac equation in one dimension 
("time"), with the field-dependent mass matrix. 
This lead to the simple procedure for finding the complete set 
of "regular" (i.e. present in the complex $\Gamma$ case) 
integrals of motion. Then an alternative but closely related 
procedure, based on the Lax pair approach, was used 
to check the completeness of the obtained set of IOM. Both procedures 
were extended to the special case $\Gamma\in{\Bbb R}$, to obtain an 
additional ("exceptional") set of IOM. 

Afterwards, the concept of symmetries of the "regular" IOM was defined 
(the I-symmetries), 
and the Lie algebras (and groups) corresponding to the linear I-symmetries 
were found. These are the $su(2)$ symmetry for the transmission gratings 
and the $su(1,1)$ symmetry for the reflection gratings. The initial doublets 
of basic fields, which were introduced as a convenience for 
more compact calculations, turned out to be the fundamental 
(i.e. spinor) representation of those symmetry algebras.
Also, the Lax matrices, constructed from these basic spinors, 
transform in the regular way, i.e. they form the adjoint representation
of the I-symmetries.

In the special case $\Gamma\in{\Bbb R}$ the number of IOM increases,
so only the subset of "regular" I-symmetries survives. This is to be expected, 
since the I-symmetries now have to satisfy a larger set of constraints 
than in the general ($\Gamma\in{\Bbb C}$) case.  

In the second part of the paper another type of symmetries was
considered, the symmetries of EOM (the E-symmetries).
The corresponding symmetry algebras are the products of several abelian 
factors (one noncompact $\sim {\Bbb R}^1$, and two compact $\sim u(1)$) 
and of one $su(1,1)$ factor (for both geometries). The action of
these symmetries on the regular IOM  was studied and a special 
kind of N\"other theorem is found to be valid here. 

In the special case $\Gamma\in{\Bbb R}$ the number of independent 
EOM gets smaller, leading to the increase in the number of E-symmetries.
This is clearly the opposite behavior to the case of I-symmetries. 
Further study is necessary to clarify the 
relation between the "regular" and "exceptional" cases.
At the end of this part, the action of the E-symmetries on the
I-symmetries was considered (in the "regular" case). 
The non-abelian factor of E-symmetries commutes with the I-symmetries,
and the two $u(1)$ factors act as rotations in the $1-2$ plane of 
I-symmetries.  

As a short excursion from the algebraic orientation of the paper, 
Section  \ref{section_on_sols} is devoted to the solutions of EOM
in two "exceptional" cases: $\Gamma\in{\Bbb R}$ and $\Gamma\in i{\Bbb R}$.
In both cases it is relatively straightforward to obtain the general
solutions (two methods for $\Gamma\in{\Bbb R}$ were presented and one 
for $\Gamma\in i{\Bbb R}$), but 
satisfying the boundary conditions characteristic of 
4WM geometries required more attention.

In the last Section one possible approach to the Hamiltonian formulation of
the 4WM system was discussed. The problems that occurred in that program
were two-fold: the non-uniqueness of the choice of the Hamiltonian (Hamilton's function), 
and the non-closeness of the field-strength 2-form. 
The first problem leads to the recognition of the multi-Hamiltonian nature of 
the 4WM system, and is not really a problem. It is just the type of the 
"gauge-symmetry" of EOM. The second problem, however, is the real obstacle
to the fulfillment of the program. The structure of this obstacle is 
topological (the violation of the Jacobi and Bianchi identities). This was 
studied for one specific 
"gauge" (the choice of the Hamiltonian), and a special circumstance when this 
obstacle can be removed was found, essentially corresponding to the
$\Gamma\in{\Bbb R}$ case. 

The same $\Gamma\in{\Bbb R}$ case was then 
treated in a different way, leading to the discovery of even bigger space 
of possible Hamiltonians. The topological obstacle is 
absent in this case, and nothing prevents a full consistent application of
the Hamilton formalism, and identification of the 
corresponding Hamiltonian action of the system. 

{\bf Future work:}
The presented work contains several topics that deserve future attention.

Questions pertaining to the general class of dynamic systems:
Clarifying the freedom of choice of Hamiltonian function among IOM;  
Studying properties of E-symmetries upon the local 
(i.e. $x$-dependent) scalings of the dynamic vector field $\vec{F}$;
Finding classes of equivalency of the symplectic form $\hat{J}$ (under 
such scalings) that have the same structure of the tensor of
non-Jacobianity $\omega$; etc.

Questions related to the 4WM system in particular: Full relation
between the $\Gamma\in{\Bbb R}$ and $\Gamma\in{\Bbb C}$ E-symmetries;
Explicit resolution of boundary conditions in $\Gamma\in{\Bbb C}$
case; Extending the theory to {\it multiple} gratings;
etc.

The last question is particularly intriguing. Even in the case of 
single gratings, 4WM EOM possess rich 
algebraic structure. However, the writing of gratings in a 
photorefractive crystal is a dynamic holographic process, 
and more than one grating can coexist simultaneously in the same region 
of the crystal. EOM then contain terms coming from different
types of gratings, and the analysis should be much more involved.

{\bf Acknowledgements:} One of the authors (PLS) expresses gratitude 
to the Brown University for support during the graduate years, when
some of the ideas explored in this work were conceived and 
partially developed.

\begin{appendix}

\section{The $\sigma$-metric elementary functions.}
\label{app_sigma}
\desnoime{app\_sigma.tex}

The elementary definitions and relations of 
the $c$ and $s$ functions are listed in this Appendix:
\vspace{0ex}
\ofa
c(\sigma,x) 
&:\equiv& 
\cos(\sqrt{\sigma}x) =
\left\{ 
\matrix{
\cos(g\alpha), \cr
\ch(\alpha),  
}\right.
\cr
s(\sigma,x) 
&:\equiv& 
\sin(\sqrt{\sigma}x) / \sqrt{\sigma} =
\left\{ 
\matrix{
\sin(\alpha), \cr
\sh(\alpha),  
}\right.
\cr 
&&
\zffa{uopstenefunkcije}

\ni where the upper/lower option 
corespond to $\sigma=+1/-1$ signs.  
\vspace{0ex}
\of
c(\sigma,x)^2 + \sigma s(\sigma,x)^2 =1  ,
\zf
\vspace{-1ex}
\ofa
2c(\sigma,x)s(\sigma,x) &=& s(\sigma,2x)               , \cr
2c(\sigma,x)^2   &=& 1+c(\sigma,2x)             , \cr
2\sigma s(\sigma,x)^2 &=& 1-c(\sigma,2x)        , \cr
c(\sigma,x)^2 - \sigma s(\sigma,x)^2 &=& c(\sigma,2x)  ,
\zfa
\vspace{-3ex}
\ofa
t(\sigma,x) &:\equiv& {s(\sigma,x)}/{c(\sigma,x)}  , \cr
ct(\sigma,x) &:\equiv& 1/{t(\sigma,x)} ,
\zffa{fje}
\vspace{-3ex}
\ofa
t(\sigma,x) + \sigma ct(\sigma,x) &=& {2\sigma}/{s(\sigma,2x)}  ,  \cr
t(\sigma,x) - \sigma ct(\sigma,x) &=& -2\sigma ct(\sigma,2x) ,
\zfa
\vspace{-3ex}
\ofa
c'(\sigma,x) &=& - \sigma s(\sigma,x)  , \cr
s'(\sigma,x) &=& c(\sigma,x) .
\zfa

\section{An alternative Laxian approach}

\label{app_lax}

\desnoime{\it app\_lax.tex}

In order to achieve a sufficient degree of generality, 
one should use the "big spinor" 
\vspace{0ex}
\of 
\left|\Psi\right>  :\equiv \pmatrix{B_1\cr \alpha B_3\cr \beta B_4\cr \gamma B_2},
\zf

\ni with arbitrary complex numbers $\alpha$, $\beta$ and $\gamma$.
Its evolution equation is
\vspace{0ex}
\of
\partial_z \left|\Psi\right> = {\bf \cal N}\left|\Psi\right> ,
\zf

\ni where 
\vspace{0ex}
\of
{\bf \cal N}=\pmatrix{
0&\sigma\mu/\alpha&0&0\cr
-\alpha\bar{\mu}&0&0&0\cr
0&0&0&-\beta\mu/\gamma\cr
0&0&\sigma\gamma\bar{\mu}/\beta&0
}.
\zf

The problem is to determine the evolving member of the Lax pair.

\subsection{${\bf \cal L}\sim \left|\Psi\right>\left<\Psi\right|$}

The matrix ${\bf \cal L}$ is searched first in the form ${\bf \cal L}={\bf \cal A}\left|\Psi\right>
\left<\Psi\right| {\bf \cal B}$, where ${\bf \cal A}$ and ${\bf \cal B}$ 
are some constant matrices. Then
\vspace{0ex}
\of
\partial_z {\bf \cal L} = 
{\bf \cal A}{\bf \cal N}{\bf \cal A}^{-1}{\bf \cal L}
+
{\bf \cal L}{\bf \cal B}^{-1}{\bf \cal N}^\dagger{\bf \cal B}
.
\zf

\ni Require 
\vspace{0ex}
\of
{\bf \cal B}^{-1}{\bf \cal N}^\dagger{\bf \cal B}
=
- {\bf \cal A}{\bf \cal N}{\bf \cal A}^{-1},
\zf

\ni i.e. 
\vspace{0ex}
\of
{\bf \cal N}^\dagger{\bf \cal C} 
= - {\bf \cal C}{\bf \cal N},
\zf
\ni where ${\bf \cal C}:\equiv {\bf \cal B}{\bf \cal A}$.

The solution of this equations is
\vspace{0ex}
\of
{\bf \cal C} = \pmatrix{
\bar{\alpha} \xi_2&
\bar{\alpha}\mu \xi_1&
\bar{\alpha} \xi_4&
\bar{\alpha}\mu \xi_3 \cr
-\alpha\bar{\mu} \xi_1&
\frac\sigma\alpha \xi_2& 
\frac{\sigma\gamma\bar{\mu}}{\beta} \xi_3&
-\frac{\beta}{\gamma} \xi_4\cr
\frac{\sigma\bar{\gamma}}{\bar{\beta}} \xi_6&
\frac{\sigma\bar{\gamma}\mu}{\bar{\beta}} \xi_5&
\frac{\sigma\bar{\gamma}}{\bar{\beta}} \xi_8&
\frac{\sigma\bar{\gamma}\mu}{\bar{\beta}} \xi_7\cr
\alpha\bar{\mu} \xi_5&
-\frac{\sigma}{\alpha} \xi_6&
-\frac{\sigma\gamma\bar{\mu}}{\beta} \xi_7&
-\frac{\beta}{\gamma} \xi_8
}.
\zf

In the general case of complex $\Gamma$ the 
factor $\mu$ is non-constant (with respect to $z$), 
and some of the parameters $\xi$ have to be set to zero: $\xi_1=\xi_3=\xi_5=\xi_7=0$.  
However, if $\Gamma$ is real, the phase factor 
$\nu{:\equiv \exp({i\hbox{arg}\mu})}$
is constant, and one may redefine the time variable 
$z\rightarrow \theta$, to absorb the non-constant absolute 
value $|\mu(z)|$. In effect this permits the full set of non-zero $\xi$ 
in the above matrix ${\bf \cal C}$ (with the replacement $\mu\rightarrow\nu$).

Let us consider the general case. The presence of four non-zero $\xi$ parameters
indicates the existence of four IOM:
\vspace{0ex}
\ofa
IOM_{\xi_2}&=& \bar{\alpha} (I_1+\sigma I_3) = \bar{\alpha} q_1, 
\cr
IOM_{\xi_8}&=& \beta \bar{\gamma} (I_2+\sigma I_4) = \beta \bar{\gamma}  q_2, 
\cr
IOM_{\xi_4}&=& \bar{\alpha}\beta (\bar{B}_1 B_4-\bar{B}_3 B_2) = \bar{\alpha}\beta \bar{q}_3, 
\cr
IOM_{\xi_6}&=& \sigma\bar{\gamma} (\bar{B}_4 B_1-\bar{B}_2 B_3) = \sigma\bar{\gamma} q_3, 
\zfa

\ni where  
\vspace{0ex}
\of
IOM_{\xi_i} :\equiv \tr{\frac{\partial{\bf \cal L}}{\partial \xi_i}} 
=
\left<\Psi\left|\frac{\partial{\bf\cal C}}{\partial \xi_i}\right|\Psi\right>.
\zf

\ni These integrals are the already known "regular" IOM. Only $q_4$ is not obtained in this way.
It will be obtained in the next subsection, with a different choice of ${\bf \cal L}$.

In the $\Gamma\in{\Bbb R}$ case, there are eight free parameters ($\xi$) and there should be eight 
IOM. The first four of them are the "regular" ones $\{q_1,q_2,q_3,\bar{q}_3\}$, and the
additional four are:
\vspace{-1ex}
\ofa
IOM_{\xi_1}&=& 
|\alpha|^2 (\nu \bar{B}_1B_3-\bar{\nu}\bar{B}_1B_3) = |\alpha|^2 w_1, 
\cr
IOM_{\xi_7}&=& 
\sigma |\gamma|^2 (\nu \bar{B}_4B_2-\bar{\nu} B_4\bar{B}_2) = - |\gamma|^2 w_2, 
\cr
IOM_{\xi_3}&=& 
\bar{\alpha}\gamma (\nu\bar{B}_1 B_2+\sigma\bar{\nu}\bar{B}_3 B_4) = -\sigma \bar{\alpha}\gamma w_3, 
\cr
IOM_{\xi_5}&=& 
 -\sigma \alpha \bar{\gamma}\bar{w}_3.
\cr 
&&
\zfa
\vspace{0ex}
\ni Here, again, all obtained integrals are already known. They are the elements of the 
$w$-set derived in the spinorial approach. The remaining elements of that set will 
be obtained in the next subsection.
\subsection{${\bf \cal L}\sim \left|\Psi\right>\left<\bar{\Psi}\right|$}

Now search for ${\bf \cal L}$ in the form ${\bf \cal L}={\bf \cal A}\left|\bar{\Psi}\right>
\left<\Psi\right| {\bf \cal B}$, where ${\bf \cal A}$ and ${\bf \cal B}$ 
are some constant matrices (different from the ones in the previous subsection). Then
\vspace{0ex}
\of
\partial_z {\bf \cal L} = 
{\bf \cal A}{\bf \cal N}{\bf \cal A}^{-1}{\bf \cal L}
+
{\bf \cal L}{\bf \cal B}^{-1}{\bf \cal N}^T{\bf \cal B}
,
\zf

\ni and the requirement that ${\bf \cal L}$ satisfies the Lax-type  evolution 
equation has the form:
\vspace{0ex}
\of
{\bf \cal B}^{-1}{\bf \cal N}^T{\bf \cal B}
=
- {\bf \cal A}{\bf \cal N}{\bf \cal A}^{-1},
\zf

\ni i.e. 
\vspace{0ex}
\of
{\bf \cal N}^T{\bf \cal C} 
= -{\bf \cal C}{\bf \cal N},
\zf

\ni where, again,  ${\bf \cal C}:\equiv {\bf \cal B}{\bf \cal A}$.
The solution to this condition is
\vspace{0ex}
\of
{\bf \cal C} = \pmatrix{
\alpha\bar{\mu} \xi_2&
\xi_1&
\alpha\bar{\mu} \xi_8&
\alpha \xi_5 \cr
-\xi_1&
\frac{\sigma\mu}{\alpha} \xi_2& 
\frac{\sigma\gamma}{\beta} \xi_5&
-\frac{\beta\mu}{\gamma} \xi_8\cr
\frac{\sigma\bar{\mu}}{\beta} \xi_7&
\frac{\sigma\gamma}{\beta} \xi_6&
\frac{\sigma\gamma\bar{\mu}}{\beta} \xi_4&
\xi_3\cr
\alpha \xi_6&
-\frac{\mu}{\alpha} \xi_7&
-\xi_3&
-\frac{\beta\mu}{\gamma} \xi_4
}.
\zf

In the $\Gamma\in{\Bbb C}$ case one has to set $\xi_2=\xi_8=\xi_7=\xi_4=0$
and the remaining $\xi$ parameters give rise to the following four "regular" IOM:
\vspace{0ex}
\ofa
IOM_{\xi_1} &\equiv& 0, \cr
IOM_{\xi_3} &\equiv& 0, \cr
IOM_{\xi_5} &=& 
\alpha\gamma (B_1B_2+\sigma B_3B_4) = \alpha\gamma q_4, \cr
IOM_{\xi_6} &=&  \alpha\gamma q_4.
\zfa

The $\Gamma\in{\Bbb R}$ case has additional $w$ integrals:
\vspace{-1ex}
\ofa
IOM_{\xi_2} &=& 
\alpha (\bar{\nu}B_1^2+\sigma\nu B_3^2)= \alpha w_4, \cr
IOM_{\xi_4} &=& 
\beta\gamma (\sigma\bar{\nu}B_4^2+\nu B_2^2)= \beta\gamma w_6, \cr
IOM_{\xi_8} &=& 
\alpha\beta (\bar{\nu}B_1B_4-\nu B_3B_2) = \alpha\beta w_5, \cr
IOM_{\xi_7} &=&  \gamma w_5.
\zfa
\ni In this way, the full sets of "regular" ($q$-set) and 
"exceptional" ($w$-set) IOMs are reconstructed. 
The presence of the arbitrary complex constants $\alpha$, $\beta$ and $\gamma$ in 
the procedure indicates that there are no
additional IOM of the bilinear type.

\section{Hyperboloids in ${\Bbb R}^4$}

\label{appendix_on_hyperboloids}

\desnoime{app\_hyperboloids.tex}

In ${\Bbb R}^4$ there exist four 
different types of the normalized
"hyperboloids", defined by $y_1^2+\epsilon_2 y_2^2 + 
\epsilon_3 y_3^2 + \epsilon_4 y_4^2 = 1$: 
\vspace{0ex}
$$
\vbox{
\offinterlineskip
\halign{
\strut  \hfil \hfil \ $#$ \ \hfil \vrule & 
\vrule\hfil\hfil  \ $#$ \ \hfil \vrule &
\vrule \hfil \ $#$ \ \hfil \vrule &
\vrule\hfil \ $#$ \  \hfil  \cr
       &\epsilon_2&\epsilon_3&\epsilon_4 \cr
\noalign{\hrule}
{\Bbb H}^3_{(4,0)} &+1&+1&+1 \cr
{\Bbb H}^3_{(3,1)} &+1&+1&-1 \cr
{\Bbb H}^3_{(2,2)} &+1&-1&-1 \cr
{\Bbb H}^3_{(1,3)} &-1&-1&-1 \cr
}
}
$$

\noindent ${\Bbb H}^3_{(4,0)}$ is just another 
name for the sphere ${\Bbb S}^3$, and ${\Bbb H}^3_{(2,2)}$
is the hyperboloid ${\Bbb H}^3$ relevant for this work.

\section{Ordered Exponential}

\label{section_on_OE_C}

\desnoime{app\_OE\_C.tex}

In this appendix some general properties of the matrix ${\bf U}(z)$ 
are discussed.

The basic 4WM EOM \jna{dirac} has a formal solution 
\vspace{0ex}
\of
\left|\psi_i(z)\right> = {\bf U}(z) \left|\psi_i(0)\right>,
\zf

\ni where ${\bf U}(z)$ satisfies the initial value problem 
\vspace{0ex}
\ofa
\partial_z {\bf U}(z) &=& {\bf m}(z) {\bf U}(z), \cr
{\bf U}(0) &=& {\bf 1}.
\zfa

\ni One can write the formal solution to this equation in the form
\vspace{0ex}
\of
{\bf U}(z) :\equiv \left(\exp\left({\int_0^z dz' {\bf m}(z')}\right)\right)_+
\zf

\ni which is called {\it the (Path) Ordered Exponential} (OE). 
The notion of ordering is referring here to 
{\bf the right-to-left} multiplication 
of the factors in the definition of OE:
\vspace{0ex}
\of
 {\left( e^{\left(\int_0^z dz' {\bf m}(z')\right)}\right)}_+ 
:\equiv 
\lim_{N\rightarrow  \infty}
\prod_{\alpha=N}^{0}
e^{\left( \frac{z}{N} {\bf m}\left(\alpha \frac{z}{N}\right)\right)},
\zf

\ni i.e. one alternates the infinitesimal integrations 
(along the path between the $z'=0$ and $z'=z$) and exponentiations
of, in such a way obtained, infinitesimal matrices.
 
OE is an entirely different object from the
ordinary matrix exponential $\exp({\int_0^z dz' {\bf m}(z')})$, 
where the whole integration along the path is performed first, 
and then only one exponentiation executed on this integrated matrix.
The source of the difference is in the non-commutativity of the 
matrices ${\bf m}(z)$ evaluated at different points.

The methods to evaluate OE are frequently non-exact: one may easily 
prove that the knowledge of ${\bf U}(z)$ for arbitrary ${\bf m}(z)$ is 
equivalent to the knowledge of the solution to the Schr\"odinger equation
for an arbitrary complex potential (and this is known to be a non-solvable 
problem).
However, in the cases when ${\bf m}(z)$ has one of the several special forms, the
exact solution for ${\bf U}(z)$ can be found. Two such cases are
encountered in this work: 

$\bullet$ for $\Gamma \in {\Bbb R}$ the matrix ${\bf m}(z)$ is proportional
to the constant matrix 
${\bf \tilda{m}}(z)=\pmatrix{0&\sigma \nu\cr -\bar{\nu}&0}$, and 
all commutators $[{\bf m}(z),{\bf m}(z')]$ are equal to zero. Thus, OE 
reduces to the ordinary exponential, and the result is displayed in 
Eq. \jna{reseno}.  

$\bullet$ for $\Gamma \in i{\Bbb R}$ the matrix ${\bf m}(z)$ has 
the raising and the lowering components that oscillate with the opposite
frequencies $\Omega$. The Appendix \ref{section_on_solving_iR_OE} gives 
one possible way to obtain ${\bf U}(z)$ for such an ${\bf m}(z)$.

In the general case, the OE ${\bf U}(z)$ satisfies several simple identities,
induced by the properties of ${\bf m}(z)$:

$\bullet$ From the tracelessness of ${\bf m}(z)$ follows the
unimodality condition $\det{\bf U}(z)=1$ ($\forall z$):
\vspace{-1ex}
\ofa
\det{\bf U} &=&
\lim_{N\rightarrow  \infty}
\prod_{\alpha=N}^{0}
\det \left[
\exp\left({ \frac{z}{N} {\bf m}\left(\alpha \frac{z}{N}\right)}\right)
\right] =\cr
&=& 
\lim_{N\rightarrow  \infty}
\prod_{\alpha=N}^{0}
\exp\left({ \frac{z}{N} \tr{\bf m}\left(\alpha \frac{z}{N}\right)}\right)
=\cr
&=& 
\lim_{N\rightarrow  \infty}
\prod_{\alpha=N}^{0}
\exp({0}) = 1.
\zfa

$\bullet$ From the membership of ${\bf m}(z)$ in the Lie algebra 
$g :\equiv su\left(({3+\sigma})/2,({1-\sigma})/2\right)$ it follows that ${\bf U}(z)$
is an element of the corresponding Lie group
${\cal G}:\equiv SU\left(({3+\sigma})/2,({1-\sigma})/2\right)$:
\vspace{0ex}
\of
{\bf U}(z)^\dagger {\bf \eta} {\bf U}(z) = {\bf \eta},
\zf

\ni where the ${\bf \eta}={\bf n}_1 =\pmatrix{1&0\cr0&\sigma}$ is the 
metric matrix of the algebra $g$. To check this identity, one should follow the 
chain of arguments:
\vspace{0ex}
\ofa
{\bf m}\left(\alpha\frac{z}{N}\right) \in g 
&\Rightarrow&
\frac{z}{N}{\bf m}\left(\alpha\frac{z}{N}\right) \in g 
\cr
&\Rightarrow&
\exp\left({\frac{z}{N}{\bf m}\left(\alpha\frac{z}{N}\right)}\right) \in {\cal G}
\cr
&\Rightarrow&
\prod_\alpha \exp\left({\frac{z}{N}{\bf m}\left(\alpha\frac{z}{N}\right)}\right) \in {\cal G}
\cr
&\Rightarrow&
{\bf U}(z) \in {\cal G}.
\zfa

\section{Ordered Exponential Solution for $\Gamma\in i{\Bbb R}$}

\label{section_on_solving_iR_OE}

\desnoime{app\_OE\_iR.tex}

In this appendix the initial value problem  
\vspace{-1ex}
\ofa
\partial_z {\bf U}(z) &=& {\bf m}(z) {\bf U}(z), \cr
{\bf U}(0) &=& {\bf 1},
\zfa

\ni is solved for $\Gamma$ imaginary. Starting with the ansatz
\vspace{0ex}
\of
U_{ij}(z) =  u_{11} \exp({i\omega_{ij} z}),
\zf

\ni one obtains a set of conditions
\vspace{0ex}
\ofa
\omega_{21} &=& \omega_{11}+\Omega, \cr
\omega_{12} &=& \omega_{22}-\Omega, \cr
\frac{u_{21}}{u_{11}} 
&=& \frac{i\omega_{11}}{\sigma\mu_0}
=  \frac{-\bar{\mu}_0}{i\omega_{21}}, \cr
\frac{u_{12}}{u_{22}} 
&=& \frac{i\omega_{22}}{-\bar{\mu}_0}
=  \frac{\sigma\mu_0}{i\omega_{12}}. 
\zfa

\ni The second pair of these conditions defines the consistency conditions
\vspace{0ex}
\of
\omega_{11}\omega_{21} = \omega_{12}\omega_{22} = \sigma |\mu_0|^2,
\zf

\ni which are converted into auxiliary equations:
\vspace{0ex}
\ofa
\omega_{11}^2 + \Omega \omega_{11} -\sigma |\mu_0|^2 &=& 0, \cr
\omega_{22}^2 - \Omega \omega_{22} -\sigma |\mu_0|^2 &=& 0.
\zfa

\ni Solutions to these quadratic equations are
\vspace{0ex}
\ofa
\omega_{11\pm} = \omega_{12\pm} &=&  (-\Omega \pm \Xi)/2, \cr
\omega_{21\pm} = \omega_{22\pm} &=& (+\Omega \pm \Xi)/2,
\zfa

\ni where $\Xi :\equiv {\tilda{\Gamma} \sqrt{q}}/{I}$ and 
$q:\equiv q_5^2+4\sigma|Q|^2$. Then 
\vspace{0ex}
\of
U_{ij} = u_{ij+} \exp({iz\omega_{ij+}}) + u_{ij-} \exp({iz\omega_{ij-}}),
\zf

\ni and one has to solve the remaining conditions 
\vspace{0ex}
\ofa
\frac{u_{21\pm}}{u_{11\pm}} &=& \frac{i \sigma \omega_{11\pm}}{\mu_0},\cr
\frac{u_{12\pm}}{u_{22\pm}} &=& \frac{- i\omega_{22\pm}}{\bar{\mu}_0},
\zfa

\ni in conjunction with the initial conditions
\vspace{0ex}
\ofa
u_{11+}+u_{11-} &=& 1,\cr
u_{12+}+u_{12-} &=& 0,\cr
u_{21+}+u_{21-} &=& 0,\cr
u_{22+}+u_{22-} &=& 1. 
\zfa 

The solution is 
\vspace{0ex}
\ofa
u_{11+} = - \frac{\omega_{11-}}{\Sigma},
& &
u_{11-} = + \frac{\omega_{11+}}{\Sigma}, 
\cr
u_{22+} = - \frac{\omega_{22-}}{\Sigma},
& &
u_{22-} = + \frac{\omega_{22+}}{\Sigma}, 
\cr
u_{12+} = - \frac{i\sigma \mu_0}{\Sigma},
& &
u_{12-} = + \frac{i\sigma \mu_0}{\Sigma},
\cr
u_{21+} = + \frac{i \bar{\mu}_0}{\Sigma},
& &
u_{21-} = - \frac{i \bar{\mu}_0}{\Sigma},
\zfa

\ni leading to the final form:
\vspace{0ex}
\ofa
U_{11} 
&=&  
\exp\left({-i\frac{\Omega z}2}\right)
\left[
\cos\left(\frac{\Xi z}2\right)
+ 
i \frac{\Omega}{\Xi}
\sin\left(\frac{\Xi z}2\right)
\right],
\cr
U_{12} 
&=&  
i\frac{2\sigma Q}{\sqrt{q}} 
\exp\left({-i\frac{\Omega z}2}\right)
\sin\left(\frac{\Xi z}2\right),
\cr
U_{21} 
&=&  
-i\frac{2\sigma \bar{Q}}{\sqrt{q}} 
\exp\left({i\frac{\Omega z}2}\right)
\sin\left(\frac{\Xi z}2\right),
\cr
U_{22} 
&=&  
\exp\left({i\frac{\Omega z}2}\right)
\left[
\cos\left(\frac{\Xi z}2\right)
- 
i \frac{\Omega}{\Xi}
\sin\left(\frac{\Xi z}2\right)
\right]
.
\cr 
&&
\zfa
It is easy to check the unimodality condition $\det{\bf U}(z)=1$ 
($\forall z$).

\section{Components of $\Omega$}

\label{appendix_with_components_of_Omega}

\desnoime{app\_Omega}
 
\ni The non-vanishing components of $\Omega$ are
\vspace{0ex}
\ofa
\Omega^{B_1\bar{B}_3B_2} &=& -\bar{\mu}\bar{\partial}_4 \mu, \cr
\Omega^{B_1\bar{B}_3B_3} &=& -\bar{\mu}\bar{\partial}_1 \mu, \cr
\Omega^{B_1\bar{B}_3B_4} &=& +\bar{\mu}\bar{\partial}_2 \mu, \cr
\Omega^{B_1\bar{B}_3\bar{B}_1} &=& -\bar{\mu}\partial_3 \mu, \cr
\Omega^{B_1\bar{B}_3\bar{B}_2} &=& -\mu\partial_4 \mu, \cr
\Omega^{B_1\bar{B}_3\bar{B}_3} &=& +\mu\partial_1 \mu, \cr
\Omega^{B_1\bar{B}_3\bar{B}_4} &=& +\bar{\mu}\partial_3 \mu, \cr
\Omega^{B_2\bar{B}_4B_1} &=& -\mu\bar{\partial}_3 \bar{\mu}, \cr
\Omega^{B_2\bar{B}_4B_3} &=& +\bar{\mu}\bar{\partial}_1 \bar{\mu}, \cr
\Omega^{B_2\bar{B}_4B_4} &=& +\mu\bar{\partial}_2 \bar{\mu}, \cr
\Omega^{B_2\bar{B}_4\bar{B}_1} &=& -\bar{\mu}\partial_3 \bar{\mu}, \cr
\Omega^{B_2\bar{B}_4\bar{B}_2} &=& -\mu\partial_4 \bar{\mu}, \cr
\Omega^{B_2\bar{B}_4\bar{B}_3} &=& +\mu\partial_1 \bar{\mu}, \cr
\Omega^{B_2\bar{B}_4\bar{B}_4} &=& +\bar{\mu}\partial_2 \bar{\mu}.
\zfa

\section{Lagrange formulation of generalized Hamiltonian
Systems}

\desnoime{app\_lagrange.tex}
\label{app_lagrange}

\subsection{From the singular Lagrangian to the generalized Hamiltonian 
dynamics}

%
%

The singular Lagrangian ($\mu\in \overline{1,D}$):
\vspace{0ex}
\of 
L = {\cal A}_{\mu}(x) \dot{x}^{\mu} - V(x)  ,
\zff{lagrangean}

\noindent 
generates the Euler-Lagrange equations of motion:
\vspace{0ex}
\ofa 
\cal{E}_{\mu} 
&:\equiv& 
{\partial L \over  \partial
x^{\mu}} - {d\over dt} 
{\partial L \over \partial \dot{x}^{\mu}}
 = 
F_{\mu\nu}\dot{x}^{\nu} - \partial_{\mu}V = 0, 
\zffa{eleom}

\noindent where
$F_{\mu\nu}:\equiv
\partial_{\mu}{\cal A}_{\nu}-\partial_{\nu}{\cal A}_{\mu}$ 
is the
tensor of the "field strength". If it is nonsingular ($\det{\bf F}\ne
0$; possible only for even $D$), 
its inverse, the tensor of symplectic structure 
${\bf J} :\equiv {\bf F}^{-1}$  can be defined. 
Then Eq. \jna{eleom} have the form of the
generalized Hamilton equations:
\vspace{0ex}
\of 
\dot{x}^{\mu} = J^{\mu\nu}\partial_{\nu}V. 
\zff{hameom}

\noindent To initiate the Hamilton formulation, one starts with the 
definition of canonical momenta
\vspace{0ex}
\of
p_{\mu} :\equiv {\partial L \over \partial \dot{x}^{\mu}} 
= {\cal A}_{\mu}(x)  ,
\zff{definicijaimpulsa}

\noindent and of the (naively defined) Hamiltonian function
\vspace{0ex}
\of 
H :\equiv p_{\mu}\dot{x}^{\mu}-L . 
\zff{naivelydefinedhamiltonian}

\noindent 
The definitions of momenta \jna{definicijaimpulsa} 
are supposed to be (non-singular) contact 
transformations replacing velocities $\dot{x}^\mu$ 
by the momenta $p_\nu$, thus allowing (by inversion) 
to express the velocities $\dot{x}^{\mu}$ in terms of the 
momenta $p_{\mu}$. Here, however, these equations are singular:
the velocities do not figure (at all) on their right-hand-sides (RHS).
Thus, instead of being the (successful) contact transformations, 
these equations are the constraints on the Hamiltonian dynamics of system:
\vspace{0ex}
\of 
\kappa_{\mu} :\equiv p_{\mu} - {\cal A}_{\mu}(x). 
\zff{veze}

\noindent
Constraints obtained from the contact transformations 
are called {\bf the primary constraints}, implying that there may be some 
additional constraints in the system (all these additional constraints are 
called {\bf secondary constraints}). Hamiltonian systems with constraints
are treated by the Dirac method which outlines are given in the rest of 
this Appendix. 

For the specific system at hand, the naively defined Hamiltonian 
\jna{naivelydefinedhamiltonian} has the arbitrary weighted terms 
proportional to primary constraints:
\vspace{0ex}
\of
H = V + \kappa_\mu \dot{x}^\mu \simeq V
\zf

\noindent Its unique (non-arbitrary) part is called 
the {\bf Canonical Hamiltonian $H_c=V$}. 
For the further purposes, one needs the (temptative) 
{\bf Total Hamiltonian}:
\vspace{0ex}
\of
H'_T = H_c + \kappa_\mu v^\mu.
\zf

\noindent where $v^\mu$ are the {\bf Lagrange multipliers}, 
for this moment taken to be arbitrary functions of time $z$. 
The prime on $H'_T$ denotes the temptative nature of this quantity: 
once when (and if) the Lagrange multipliers are determined
(i.e. replaced with suitable functions over phase space), 
this quantity will be replaced by symbol $H_T$. One can not know apriory 
what values will $v$s acquire. Instead, the (Dirac's) constraint analysis 
has to be performed to determine the full set of constraints in
the system and then to classify the constraints as either {\bf the 
first class}
(ones that commute in a weak sense with all other constraints)
or {\bf the second class} (ones that are not of the first class). 
Then, the Lagrange multipliers standing next to the 
primary constraints of the second class will be determined as a specific 
functions on the phase space, while the Lagrange multipliers corresponding 
to the primary constraints of the first class will stay undetermined.
Presence of the second class constraints leads to the reduction of the 
phase space, while the first class constraint generate the gauge 
transformations (on the phase space).   

 The phase space $\Gamma=\{(x,p)|\forall x,p\in{\Bbb R}^D\}$ possesses 
the {\bf Poisson bracket}:
\vspace{0ex}
\of 
\{f,g\}_{PB} 
:\equiv 
\frac{\partial f}{\partial x^\mu}
\frac{\partial g}{\partial p_\mu}
-
\frac{\partial f}{\partial p_\mu}
\frac{\partial g}{\partial x^\mu}
.
\zf

\ni 
The evolution of any function $f$ on the phase space $\Gamma$
is defined by
\vspace{0ex}
\of
\frac{df}{dt}=\{f,H'_T\}_{PB}.
\zf

\noindent The consistency equations of the primary constraints 
\vspace{0ex}
\of
\frac{d \kappa_\mu}{d t} 
= 
\{\kappa_\mu, H'_T\}_{PB} 
= 
-\partial_\mu V + F_{\mu\nu}v^\nu =0, 
\zff{ppoorrtt} 

\noindent can be solved if ${\bf F}=(F_{\mu\nu})$ is non-singular,
giving the final expression for the Lagrange multipliers:
\vspace{0ex}
\of
v^\mu = J^{\mu\nu}\partial_\nu V,
\zf

\noindent 
If ${\bf F}$ is singular, some components of the RHS  
of Eq. \jna{ppoorrtt} can not be solved for $v$, and one has 
to define the secondary constrains (and then to check their 
consistency, and so on). In this work, only the case of 
non-singular ${\bf F}$ is discused.

Here, all primary constrains are of the second class, 
i.e. each of them has (at least) one other constraint 
that does not commute (in Poisson bracket sense) with it.  
This is easy to see from
\vspace{0ex}
\of 
\left\{\kappa_{\mu},\kappa_{\nu}\right\}_{PB} = F_{\mu\nu}(x),
\zf

\noindent and the non-singularity of the matrix
${\bf F}$. Constraints of the second class reduce
the phase space of the system (whereas  the first class
constraints, if they existed,  would be the gauge symmetries of the
corresponding action). 

To successfully perform the reduction of phase space, one needs to 
replace the Poisson brackets with the Dirac brackets, defined as:
\vspace{0ex}
\ofa 
\left\{f,g\right\}_{DB}  
&:\equiv& 
\left\{f,g\right\}_{PB} -
\left\{f,\kappa_{\mu}\right\}_{PB} 
J^{\mu\nu}
\left\{\kappa_{\nu},g\right\}_{PB}. 
\cr
&& 
\zfa

\noindent With respect to this structure, the connections $\kappa$
are constant, i.e. every function $f(x,p)$ on the phase space
commutes with them:
\vspace{0ex}
\of 
\left\{f,\kappa_{\mu}\right\}_{DB} = 0 . 
\zf

\noindent On this way, one can consistently work with the reduced phase space 
$\Gamma^* :\equiv \Gamma/\{\kappa=0\}=\{(x, p={\cal A}(x))|\forall 
x\in{\Bbb R}^D\}$. The Poisson bracket on this space 
is defined as 
\vspace{0ex}
\of
\{f,g\}_{PB \ on \ \Gamma^*} 
:\equiv 
\left.
\{f,g\}_{DB \ on \ \Gamma}
\right|_{\kappa=0}.
\zf

Since the  coordinates $x^{\mu}$ do not commute 
with respect to the Dirac bracket 
\vspace{0ex}
\of 
\left\{x^{\mu},x^{\nu}\right\}_{DB} = J^{\mu\nu},
\zff{nekomutativno}

\noindent on the space $\Gamma$, 
their Poisson bracket on $\Gamma^*$ are non-vanishing, too:
\vspace{0ex}
\of 
\left\{x^{\mu},x^{\nu}\right\}_{PB \ on \ \Gamma^*}  
= \left.J^{\mu\nu}\right|_{\Gamma^*}.
\zff{nekomutativno2}

\noindent From non-singularity of the tensor ${\bf J}$ 
 one can conclude that the half of coordinates $x$
can be used as the real coordinates on the $\Gamma^*$, 
and rest of them are the corresponding conjugated momenta.

\subsection{When the generalized 
Hamiltonian dynamics has the Lagrangian
formulation?}

One can turn any even-dimensional generalized Hamiltonian system (given
by Eqs \jna{hameom}) into the Lagrangian \jna{lagrangean} iff:

a) Matrix ${\bf J}$ is nonsingular, so one can define its inverse ${\bf F}$, and

b) 2-form $\hat{F} :\equiv {1\over 2} F_{\mu\nu}
dx^{\mu}\wedge dx^{\nu}$ is closed, i.e. satisfies the Darboux
condition $d \hat{F} = 0$.

The two conditions are (in a simple connected
region $U$ of the phase space: $\pi_1(U)=0$) 
sufficient to assure the
existence of "potentials" $\vec{\cal A} = {\cal A}_{\mu} dx^{\mu}$, 
such that ${\hat{F}}$ is exact 2-form
${\hat{F}} = d \vec{\cal A}$ (and, therefore, closed). 
Under these conditions the potentials are:
\vspace{0ex}
\of 
{\cal A}_\mu(x) = \frac12 \int_{x_0}^x  F_{\mu\nu}(y) dy^\nu , 
\zf

\noindent where the integration is along any  path connecting
points $x_0$ and $x$, which belongs to the domain $U$, and
$x_0$ is the point where ${\cal A}_\mu(x_0)=0$.

The action has the form:
\vspace{0ex}
{\small
\ofa
S[t_1,t_2] &=& 
\frac12 \int_{x_{1}}^{x_{2}} dx^{\nu}\int_{x_0}^x dy^{\mu}
F_{\mu\nu}(y) 
- \int_{t_{1}}^{t_{2}} dt H(x(t)) 
 .
\cr
&&
\zfa
}

\noindent The first term is the weighted surface integral over the
surface spanned by points $x_0$, $x_1$ and $x_2$. The second term is
the line term, i.e. it lives only on the line that connects the
points $x_1$ and $x_2$.

\end{appendix}

\end{multicols}

\end{document}